# ADIABATIC AND RADIATIVE COOLING OF RELATIVISTIC ELECTRONS APPLIED TO SYNCHROTRON SPECTRA AND LIGHT-CURVES OF GAMMA-RAY BURST PULSES

A. PANAITESCU

Space & Remote Sensing, MS D466, Los Alamos National Laboratory, Los Alamos, NM 87545, USA

## ABSTRACT

We investigate the adiabatic and radiative (synchrotron and inverse-Compton) cooling of relativistic electrons whose injected/initial distribution with energy is a power-law above a typical energy $\gamma_i$. Analytical and numerical results are presented for the *cooling-tail* and the *cooled-injected* distribution that develop below and above the typical energy of injected electrons, for the evolution of the peak-energy $E_p$ of the synchrotron emission spectrum, and for the pulse shape (rise and decay) resulting from an episode of electron injection.

The synchrotron emission calculated numerically is compared with the spectrum and shape of Gamma-Ray Burst (GRB) pulses. Both adiabatic and radiative cooling processes lead to a softening of the pulse spectrum, manifested both as a decreasing peak-energy or a softening of the low-energy photon spectrum slope $\alpha$, and both types of cooling processes lead to pulses peaking earlier and lasting shorter at higher energy, quantitatively consistent with observations.

The measured GRB low-energy spectrum slope $\alpha$ and the pulse shape (time-asymmetric, with a short rise and a long decay) can be used to constrain the histories of the electron injection rate $R_i$ and magnetic field $B$.

For *adiabatic*-dominated electron cooling, a power-law injection rate $R_i(t)$ suffices to explain the observed power-law GRB low-energy spectra. *Synchrotron*-dominated cooling leads to power-law cooling-tails that yield the synchrotron standard slope $\alpha = -3/2$ provided that $R_i \sim B^2$, which is exactly the expectation if the magnetic field is a constant fraction of the post-shock energy density. Increasing (decreasing) $R_i(t)$ and decreasing (increasing) $B(t)$ lead to slopes $\alpha$ harder (softer, respectively) than the standard value and to non–power-law (curved) cooling-tails. *Inverse-Compton* cooling yields four canonical values for the slope $\alpha$ but, as for synchrotron, other $R_i$ or $B$ histories yield a wider range of slopes and curved low-energy spectra. Feedback between the power-law segments that develop below and above the typical injected electron leads to a synchrotron spectrum with many breaks above and below the usual 10 keV–1 MeV observing range.

Because adiabatic cooling becomes dominant at smaller source radii, where the pulse is likely to be shorter, we expect signatures of adiabatic cooling (smaller fractional peak lag, pulses being more time-symmetric with increasing photon energy) to be seen more often in shorter pulses/bursts.

The width of the synchrotron spectrum $\varepsilon F_\varepsilon$ is determined primarily by the low ($\alpha$) and high-energy ($\beta$) photon slopes and by the width of the synchrotron function (which can be well approximated by t he Band function with $\alpha = -2/3$ and an arbitrarily large $\beta$), with a negligible contribution from the integration over the curved emitting surface, and without any broadening from cooling at the injected electron energy $\gamma_i$.

## 1. INTRODUCTION

In this work, we attempt to provide a rigurous and comprehensive **analytical** treatment of adiabatic and radiative cooling of relativistic electrons with an initial power-law distribution with energy above a constant typical energy $\gamma_i$, as expected for first-order Fermi acceleration at relativistic shocks. That is achieved by identifying *power-law* solutions to the equation for conservation of particles, given the corresponding electron cooling-law (adiabatic, synchrotron, or inverse-Compton).

For an injected power-law electron distribution with energy and a particle cooling rate that is a power-law of the particle energy $\gamma$, the resulting *cooled-injected* distribution at energy $\gamma > \gamma_i$ is a power-law or, temporarily, a broken power-law. However, the cooling-tail that develops at $\gamma < \gamma_i$ is a power-law only under certain conditions for the only two factors at play: the particle injection rate $R_i(t)$ and the magnetic field $B(t)$. Evidently, for adiabatic cooling, the effective particle distribution is set only by $R_i$. For synchrotron cooling, $B(t)$ alone sets the electron cooling, while inverse-Compton cooling is determined by both $B(t)$ and $R_i(t)$ because both quantities set the energy density of the seed synchrotron photons. The latter limits how accurate the analytical treatment of inverse-Compton cooling can be and motivates a **numerical** treatment for that case, which is also required to calculate non–power-law cooling-tails that develop for more general injection rate and magnetic field histories.

We also determine *analytically* the evolution of the peak energy $E_p$ of the power-per-decade $\varepsilon F_\varepsilon$ instantaneous spectrum and the pulse shape/light-curve $C_\varepsilon$ for the **synchrotron** emission from the electron distribution resulting from an episode of injection in the shock down-stream region, undergoing adiabatic or synchrotron cooling, and compare *numerically*-calculated synchrotron spectra and light-curves with those measured for GRB pulses.

There are many temporal/spectral features and spectral-temporal correlations of GRB pulses, identified mostly from CGRO-BATSE observations; we will contrast the synchrotron-emission features arising from adiabatic and synchrotron-cooled electrons with the following observational features:
$i)$ the distribution of the low-energy (below $E_p$) spectral slope $\alpha$ (Preece et al 2000)
$ii)$ the progressive softening of the GRB pulse spectrum (Bhat et al 1994, Ford et al 1995, Band 1997)
$ii)$ the fast rise–slow decay temporal asymmetry of GRB pulses (Nemiroff et al 1994, Norris et al 1996, Lee, Bloom & Petrosian 2000)
$iii)$ pulse-peaks occur earlier at higher energy (Norris et al 1996)
$iv)$ pulses are shorter at higher energy (Link, Epstein & Priedhorsky 1993, Fenimore et al 1995, Norris et al 1996, Lee et al 2000)
$v)$ pulses time-asymmetry is independent of observing energy (Norris et al 1996, Lee et al 2000).

The measured low-energy spectral slope $\alpha$ and the pulse time-asymmetry factor will be used to contrain the $R_i(t)$ and $B(t)$ histories. No specific model for GRBs (central engine, internal shocks, photospheric emission) is used, except that the GRB emission is synchrotron (as advocated by Meszaros & Rees 1993, Papathanassiou & Meszaros 1996, Sari, Narayan & Piran 1996, Tavani 1996, Panaitescu & Meszaros 1998) from shock-accelerated electrons (e.g. Rees & Mészáros 1994).

## 2. ELECTRON COOLING

The evolution of electron distribution with energy, $N(\gamma) = dn/d\gamma$ arises from conservation of particles, and is given by

$$\frac{\partial N}{\partial t} = N_{inj} - \frac{\partial}{\partial \gamma}\left(N \frac{d\gamma}{dt}\right) \tag{1}$$

where $t$ is the co-moving frame time measured since the beginning of electron injection, $N_{inj}$ is the power-law distribution of electrons injected by the shock (through first order Fermi mechanism)

$$N_{inj} = \begin{cases} 0 & \gamma < \gamma_i \\ k_i \gamma^{-p} & \gamma_i < \gamma \end{cases} \tag{2}$$

and

$$-\frac{d\gamma}{dt} = \frac{P(\gamma)}{mc^2} = K \frac{\gamma^n}{(t+t_o)^x} \tag{3}$$

is the electron cooling law, with $P$ the cooling power, $K$ a constant corresponding to the dominant cooling process (adiabatic, synchrotron, or inverse-Compton), with all the time-dependencies separated as a power-law of time, and $t_o$ is the age of the shock when electron injection began. For synchrotron emission, the right-hand side of equation (3) depends on the magnetic field $B(t)$, while for inverse-Compton emission, it depends additionally on the electron optical thickness $\tau(t)$ and, possibly, on the lowest electron energy $\gamma_m(t)$ reached by $\gamma_i$ electrons after cooling. In general, the cooling-law (equation 3) does not have to be a power-law in time, but we employ a power-law for analytical integration, leaving other cases for the numerical treatment.

If the injected electrons have a power-law distribution with energy, then we search for solutions that are also power-laws (and which account for the observed power-law GRB spectra, although an electron distribution slightly curved over a factor 3 in electron energy will yield a curved synchrotron spectrum over a factor 10 in photon energy that could resemble a power-law measured spectrum):

$$N(\gamma, t) = \begin{cases} a(t)\gamma^{-m} & \gamma < \gamma_i \\ A(t)\gamma^{-q} & \gamma_i < \gamma \end{cases} \tag{4}$$

### 2.1. Cooling-tail: $\gamma < \gamma_i$

At $\gamma < \gamma_i$, where $N_{inj} = 0$, substitution of equations (3) and (4) in (1) leads to

$$\frac{da}{dt} = (n-m)\frac{aK}{(t+t_o)^x}\gamma^{n-1} \tag{5}$$

which must be satisfied for any $\gamma$, if a power-law solution exists. At a given an electron energy $\gamma < \gamma_i$, we have $a(t) = 0$ until the $\gamma_i$ electrons cool to $\gamma$, and equation (5) is trivially satisfied. The time to cool from $\gamma_i$ to $\gamma$ can be found by integrating the cooling-law given of equation (3). After that time, $a > 0$ and there are two cases:

#### 2.1.1. *n=1*

For n=1 (as for adiabatic cooling and for a particular case of inverse-Compton cooling dominated by scatterings at the Thomson–Klein-Nishina transition), the right side of equation (5) is independent of $\gamma$, just as $a(t)$, and that equation leads to $da/a = (1-m)K dt/(t+t_o)^x$, i.e. to a constant $a(t)$ if $m = 1$, to an exponential $a(t)$ if $m \neq 1$ and $x \neq 1$, or to a power-law $a(t)$ if $m \neq 1$ and $x = 1$. At this point, the index $m$ of the cooling-tail cannot be determined.

#### 2.1.2. $n \neq 1$

For n≠1 (as for synchrotron and inverse-Compton cooling), the right side of equation (5) is dependent on $\gamma$, unlike $a(t)$, which requires that $m = n$, thus, if a power-law cooling-tail develops, then that **power-law cooling-tail has an index equal to the exponent of the cooling power** $P(\gamma) \sim \gamma^n$ of equation (3). For $m = n$, equation (5) implies that $da/dt = 0$, thus *a power-law cooling-tail requires that $a$ is constant.* In other words, $a = const$ is a necessary condition (but not sufficient) for a power-law cooling-tail.

That constant $a$ can be determined from the continuity of $N(\gamma)$ at $\gamma_i$:

$$a\gamma_i^{-n} \simeq \int_{t-t_{ci}}^{t} k_i(t')dt' \gamma_i^{-p} = \frac{p-1}{\gamma_i}\int_{t-t_{ci}}^{t} R_i(t')dt' \tag{6}$$

where the middle term gives the cumulative number of electrons injected during the last cooling timescale $t_{ci}$ of the $\gamma_i$ electrons, and the right term contains the electron injection rate

$$R_i = k_i \int_{\gamma_i}^{\infty} \gamma^{-p} d\gamma = \frac{k_i}{(p-1)\gamma_i^{p-1}} \tag{7}$$

For electrons cooling at the power $d\gamma/dt$ given in equation (3), the timescale $t_{ci}$ during which the $\gamma_i$ electrons lose most of their energy is

$$t_{ci}(t) = \frac{\gamma_i}{-\frac{d\gamma}{dt}(\gamma_i)} = \frac{(t+t_o)^x}{K\gamma_i^{n-1}} \quad (8)$$

For a slowly varying injection rate (or for one that is a power-law in time and in the $t_{ci} \ll t$ limit), the integral in equation (6) is approximately $R_i t_{ci}$ and equation (6) leads to $a(t) \simeq (p-1)\gamma_i^{n-1} R_i t_{ci}$, thus $a = const$ for $R_i(t) \sim 1/t_{ci}(t)$. Therefore, **a power-law cooling-tail**

$$N(\gamma < \gamma_i) \simeq (p-1)\frac{R_i t_{ci}}{\gamma_i}\left(\frac{\gamma}{\gamma_i}\right)^{-n} \quad (9)$$

**requires**

$$R_i \sim \frac{1}{t_{ci}} \sim \frac{K}{t^x} \sim \frac{P(\gamma)}{\gamma^n} \quad (10)$$

Note that $a = const$ means that the power-law cooling-tail has a constant normalization.

### 2.2. **Cooled injected electrons:** $\gamma_i < \gamma$

Substituting the cooling-law of equation (3) and the power-law solution of equation (4) in the continuity equation (1) leads to

$$\frac{dA}{dt} + (q-n)\frac{AK}{(t+t_o)^x}\gamma^{n-1} = k_i \gamma^{q-p} \quad (11)$$

that must be satisfied for any $\gamma$, which allows the determination of the index $q$:

#### 2.2.1. *n=1*

The left side of equation (11) is independent of $\gamma$ and requires that $q = p$ so that the right side is also independent of $\gamma$, thus **the cooled-injected electron distribution has the same power-law index as the injected distribution** and $A(t)$ satisfies

$$\frac{dA}{dt} + \frac{(p-1)K}{(t+t_o)^x}A = k_i \quad (12)$$

The solution of this linear, first-order differential equation can be written expplicitly only if the injection function $k_i(t)$ is a power-law of certain indices, but we can approximate its solutions in asymptotic regimes.

As long as $dA/dt$ is dominant in equation (12), we have

$$A(t) = \int_0^t k_i(t')dt' \simeq \gamma_i^{p-1}\int_0^t R_i(t')dt' \equiv A_1(t) \quad (13)$$

Then, the assumption that $dA/dt$ is dominant can be cast as $k_i \gg K\bar{k}_i t/(t+t_o)^x$ with $\bar{k}_i = \int_0^t k_i(t')dt'/t$ being the average injection factor until time $t$, thus $dA/dt$ is dominant if $(t+t_o)^x > (\bar{k}_i/k_i)Kt$, which may be easily satisfied at all times if $x > 1$, but sets an upper limit on the time until when equation (13) holds if $x < 1$.

When $dA/dt$ is not dominant, equation (12) implies that

$$A(t) \simeq \frac{k_i(t+t_o)^x}{(p-1)K} \equiv A_2(t) \quad (14)$$

#### 2.2.2. $n \neq 1$

If $dA/dt$ is dominant in equation (11), then $q = p$ is required for that equation to be satisfied for any $\gamma$ and $A(t)$ is given again by equation (13). Then, the condition $dA/dt \gg AK\gamma^{n-1}/(t+t_o)^x$ leads to a constraint on the electron energy: $\gamma^{n-1} \ll (k_i/\bar{k}_i)(t+t_o)^x/(Kt)$. Defining the **cooling energy** $\gamma_c$ as that of the electrons that lose most of their energy on a timescale equal to the time $t$ since injection began

$$t = \frac{\gamma_c}{-\frac{d\gamma}{dt}(\gamma_c)} = \frac{(t+t_o)^x}{K\gamma_c^{n-1}} \quad (15)$$

that is

$$\gamma_c(t) = \left[\frac{(t+t_o)^x}{Kt}\right]^{1/(n-1)} = \gamma_i\left(\frac{t_{ci}}{t}\right)^{1/(n-1)} \quad (16)$$

it follows that **the cooled-injected electron distribution has the same index as the injected distribution** at $\gamma < \gamma_{c1} \equiv \gamma_c[k_i(t)/\bar{k}_i]^{1/(n-1)}$ if $n > 1$ and at $\gamma > \gamma_{c1}$ if $n < 1$, where $\bar{k}_i$ is the average of the injection constant until time $t$. Note that, if $k_i$ does not vary strongly, then $\bar{k}_i \simeq k_i(t)$ and $\gamma_{c1} \simeq \gamma_c$.

If the $dA/dt$ term is not dominant in equation (11), then that equation is satisfied for any $\gamma$ only if $q = p+n-1$, and $A(t)$ is that of equation (14). The condition that the $dA/dt$ is not dominant leads to a constraint on the electron energy for which that condition is satisfied. $q = p+n-1$ means that **the cooled-injected distribution is steeper or flatter than the injected distribution** at $\gamma < \gamma_{c_2} \equiv \gamma_c[d\ln k_i/d\ln t]^{1/(n-1)}$ if $n < 1$ and at $\gamma > \gamma_{c2}$ if $n > 1$, *depending on how strong is the electron cooling* (see next). Again, if $k_i$ is slowly varying, then $\gamma_{c2} \simeq \gamma_c$.

A *weak* cooling process with $P(\gamma) \sim \gamma^n$ and $n < 1$ (inverse-Compton scatterings in the KN regime), leads to a cooling timescale $t_c(\gamma) = \gamma/P(\gamma) \sim \gamma^{1-n}$ that increases with electron energy $\gamma$ and *hardens* the electron distribution at $\gamma < \gamma_c$, making it flatter than the injected one for electrons with a cooling timescale $t_c(\gamma)$ shorter than the current age $t$. A *strong* cooling process with $n > 1$ leads to a cooling timescale that decreases with electron energy and *softens* the electron distribution at $\gamma > \gamma_c$, making it steeper than the injected one for electrons with $t_c(\gamma) < t$.

Putting together these results, the cooled-injected electron distribution is:

$$N(\gamma > \gamma_i) = \begin{cases} A_2(t)\gamma^{-(p+n-1)} & \gamma < \gamma_c \; (t > t_{ci}) \\ A_1(t)\gamma^{-p} & \gamma_c < \gamma \; (any\ t) \end{cases} \quad (n < 1) \quad (17)$$

$$N(\gamma > \gamma_i) = \begin{cases} A_1(t)\gamma^{-p} & \gamma < \gamma_c \; (t < t_{ci}) \\ A_2(t)\gamma^{-(p+n-1)} & \gamma_c < \gamma \; (any\ t) \end{cases} \quad (n > 1) \quad (18)$$

For $n < 1$ (which happens only for a particular case of inverse-Compton cooling), equation (16) shows that $\gamma_c$ increases, crossing $\gamma_i$ at $t_{ci}$, therefore $N(\gamma > \gamma_i)$ of equation (17) has only the second branch at $t < t_{ci}$, and has two power-law segments at $t > t_{ci}$. For $n > 1$ (which is most often the case for radiative cooling), the opposite conclusions are reached: $\gamma_c$ decreases, thus $N(\gamma > \gamma_i)$ of equation (18) has two power-law segments at $t < t_{ci}$ and only the second branch at $t > t_{ci}$. For either case, the cooled-injected electron distribution steepens at $\gamma_c$.

Using equations (7, (8), and (14), it can be shown that, at $t > t_{ci}$, the cooling-tail of equation (9) and the $N(\gamma) = A_2\gamma^{-(p+n-1)}$ branch of the cooled-injected distribution just above $\gamma_i$ are continuous at $\gamma_i$, which indicates that approximations made in the derivation of equations (9) and (14) are not that bad.

### 2.3. **Numerical integration of electron cooling**

For a small numerical perturbation/error $\delta N$ to the solution $N_o$ of the continuity equation (1), the resulting equation for $\delta N(t)$ leads to an exponential growth $\delta N \sim \exp(c\gamma^{n-1}t^{1-x})$ for $x < 1$ in the cooling-law of equation (3) and to a saturated growth $\delta N \sim \exp(c\gamma^{n-1})$ for $x > 1$. Therefore, equation (1) is generally unstable and cannot be integrated through a simple finite differencing scheme. Numerically, for synchrotron cooling (n=2) in a constant magnetic field (x=0), we have observed that this instability starts at the highest electron energy, propagates to lower energies, and stops at the cooling energy $\gamma_c$ where the cooling timescale equals to the integration time.

Because of that instability and to ensure particle conservation (equivalent to allowing the use of a larger timestep), equation (1) is integrated numerically by tracking the flow of particles in energy, which entails subtracting from any cell the existing electrons that cool out during timestep $\delta t$ and adding to that cell the injected electrons that remain in it after $\delta t$ (i.e. excluding the injected electrons that cool out), the existing higher energy electrons that cool in, and the injected electrons of higher energy that cool in. The cooling-law of equation (3) can be integrated analytically for adiabatic and synchrotron cooling, where the constant $K$ can be written explicitly, but not for inverse-Compton cooling, where $K$ is an integral over the current electron distribution. Thus, for adiabatic and synchrotron cooling, the analytical solution to the cooling-law can be used for a more accurate tracking of particle flow, i.e. a larger integration timestep $\delta t$ can be used for a required accuracy.

For all cells where the timestep $\delta t$ is shorter than the timescale for electrons to cool through that cell, the numerical integration introduces a diffusion of electrons toward lower energies, which moves them faster than their natural cooling would do. That numerical diffusion can be reduced by decreasing the number of timesteps (i.e. using a longer $\delta t$); however, for a too large $\delta t$, the integration leads to spurious features either at low energies (for adiabatic cooling) or at high energies (for synchrotron cooling). Those features can be eliminated by using a coarser energy grid, which in turn enhances numerical diffusion. Thus, an accurate integration of the continuity equation may require an optimization of the energy and temporal discretizations, "accurate" meaning without spreading too much the cooling-tail and without producing artificial features at any energy.

## 3. ADIABATIC COOLING

For relativistic particles, when the adiabatic index is 4/3, adiabatic cooling is described by $\gamma V^{(4/3)-1} = const$, with $V \sim R^2\Delta'$ being the comoving-frame volume, $R = c(t+t_o)\Gamma$ is the source radius, the source Lorentz factor $\Gamma$ accounts for the time $t$ measured in the comoving frame, and $\Delta' = \Delta\Gamma$ is the comoving-frame source thickness. Shock hydrodynamics equations show that the *already* shocked fluid does not expand radially, being squeezed between the contact discontinuity and the shock, and that the thickness $\Delta$ of the shocked gas shell increases only due to the addition of newly shocked fluid, thus $\Delta = const$ for adiabatic losses, and so is $\Delta'$ if there is not a significant deceleration of the shocked gas ($\Gamma = const$). Therefore, $\gamma \sim R^{-2/3} \sim (t+t_o)^{-2/3}$, with $t$ being the time since electron injection began, at age $t_o$.

Thus, for adiabatic cooling, the lowest electron energy is

$$\gamma_m(t) = \gamma_i \left(1 + \frac{t}{t_o}\right)^{-2/3} \tag{19}$$

$$-\frac{d\gamma}{dt} = \frac{2}{3}\frac{\gamma}{t+t_o} \rightarrow K_{ad} = \frac{2}{3},\ n=1,\ x=1 \tag{20}$$

in equation (3) and the cooling timescale is independent of the electron energy:

$$t_c = \frac{\gamma}{-\frac{d\gamma}{dt}} = \frac{3}{2}(t+t_o) \tag{21}$$

Consequently, cooling of infinitesimal injected distributions is a translation in log space to lower energies. Above $\gamma_i$, the sum of such shifted infinitesimal populations has the same index as the injected distribution: $q = p$, as discussed in §2.2. For $x = 1$, equation (12) can be solved analytically if $k_i$ is a power-law in time, $k_i \sim (t+t_o)^y$, leading to $A(t) = [y + (2p+1)/3]^{-1}k_i(t+t_o)$, which is close to the approximate result of equation (13): $A(t) = [3/2(p-1)]k_i(t+t_o)$.

The cooling-tail below $\gamma_i$ must satisfy equation (5): $da/dt = (2/3)(1-m)a/(t+t_o)$, leading to $a(t) \sim (t+t_o)^{2(1-m)/3}$. Then, the continuity of the electron distribution at $\gamma_i$ and at any time, $a(t)\gamma_i^{-m} = A(t)\gamma_i^{-p}$, can be used to determine the power-law index $m$ for a power-law injection rate:

$$k_i \sim (t+t_o)^y \rightarrow m = -\frac{3y+1}{2} \tag{22}$$

This result holds only for $m < p$ because the addition of infinitesimal power-laws cannot lead to a distribution softer than the injected one. Conversely, if $-(3y+1)/2 > p$, then $m = p$.

The above analytical results for the cooling-tail and for the cooled-injected distribution

$$N(\gamma) \sim t^{y+1} \begin{cases} \gamma^{(3y+1)/2} & \gamma < \gamma_i \\ \gamma^{-p} & \gamma_i < \gamma \end{cases} \tag{23}$$

for an injection rate $k_i \sim t^y$ are verified in **Figure 1**, which shows the electron distribution for adiabatic cooling obtained by integrating numerically equation (1).

For a power-law electron distribution $N(\gamma) \sim \gamma^{-m}$, the synchrotron spectrum is a power-law in photon energy: $F_\varepsilon \equiv dF/d\varepsilon \sim \varepsilon^{-(m-1)/2}$. For hard distributions with an index $m > 1/3$, the synchrotron emission at a photon energy $\varepsilon$ is dominated by electrons whose synchrotron characteristic frequency is larger than $\varepsilon$, and the actual spectrum is $F_\varepsilon \sim \varepsilon^{1/3}$. Then, equation (23) implies that the low-energy slope of the GRB photon spectrum $dC/d\varepsilon \sim \varepsilon^\alpha$ at energies below the synchrotron characteristic frequency of the $\gamma_i$ electrons has

$$\alpha = \begin{cases} -\frac{2}{3} & -\frac{5}{9} < y \quad (\frac{1}{3} < m) \\ \frac{3y-1}{4} & -\frac{2p+1}{3} < y < -\frac{5}{9} \quad (p < m < \frac{1}{3}) \\ -\frac{p+1}{2} & y < -\frac{2p+1}{3} \quad (m = p) \end{cases} \tag{24}$$

Conversely, an injection rate $R_i \sim t^y$ with $y = (4\alpha+1)/3$ leads to a synchrotron emission from adiabatically-cooled electrons that has a low-energy spectral slope $\alpha$.

## 4. SYNCHROTRON COOLING

For a constant magnetic field, synchrotron cooling is characterized by

$$K_{sy}(x=0) = \frac{1}{6\pi}\frac{\sigma B^2}{mc} \ , \ n = 2 \tag{25}$$

with $\sigma$ being the electron cross-section for photon scattering. Integration of the cooling-law of equation (3) yields

$$\gamma_m(t) = \frac{\gamma_i}{1+\frac{t}{t_{ci}}} \ , \ \gamma_c(t) = \gamma_i \frac{t_{ci}}{t} \tag{26}$$

where the last result follows from equation (16), and with

$$t_{ci}(x=0) = \frac{6\pi mc^2}{\sigma B^2 \gamma_i} \tag{27}$$

from equation (8).

For any cooling process with $n > 1$, including synchrotron, equation (16) shows that the cooling energy $\gamma_c$ decreases, being above $\gamma_i$ at $t < t_{ci}$ and below $\gamma_i$ at $t > t_{ci}$. Equation (26) shows that, at $t \gg t_{ci}$, the minimum energy $\gamma_m$ is equal to the cooling energy $\gamma_c$.

$t < t_{ci}$, $\gamma_m \lesssim \gamma_i < \gamma_c$. Before $t_{ci}$, the cooling-tail is not yet developed. For $n > 1$, the radiative cooling timescale $t_c(\gamma)$ decreses with electron energy $\gamma$, thus the cooled-injected distribution steepens from a power-law of index $q = p$ below $\gamma_c$ to a power-law index $q = p + n - 1 = p + 1$ above $\gamma_c$, as in equation (18). That implies a steepening by 1/2 of the synchrotron spectrum across the cooling break, and is a standard result for synchrotron spectra from power-law electron distributions (Sari, Piran & Narayan 1998 have applied it to GRB afterglow spectra).

$t > t_{ci}$, $\gamma_m \simeq \gamma_c < \gamma_i$. After $t_{ci}$, the cooled-injected distribution has only the $\gamma > \gamma_c$ branch: $N(\gamma > \gamma_i) \sim \gamma^{-(p+1)}$. As shown in equation (9), the cooling-tail below $\gamma_i$ is a power-law of index $m = n = 2$ if the injection rate satisfies equation (10): $R_i \sim P_{sy}(\gamma)/\gamma^2 \sim B^2$. In this case[1], the low-energy GRB photon spectrum has a slope $\alpha = -3/2$. This another standard result (Cohen et al 1997 searched for GRBs with $\alpha = -3/2$) that is usually identified for a constant $R_i$ and $B$, but holds for any $R_i \sim B^2$.

For $R_i \sim t^y$, the (intuitive) conclusions that an increasing injection rate $R_i$ yields a harder cooled-tail and a harder GRB low-energy spectrum of index $\alpha > -3/2$, while a decreasing $R_i$ leads to a softer cooling-tail and a softer GRB spectrum of $\alpha < -3/2$, are verified in **Figure 2** (left panels). Numerically, we also find that, for a constant injection rate $R_i$, a decreasing magnetic field yields a harder cooling-tail while an increasing rate $R_i$ softens that tail (Figure 2, right panels). If the injection rate and the magnetic field evolve in same direction, their effect on the index of the cooling-tail compensate each other.

## 5. INVERSE-COMPTON COOLING

The inverse-Compton (iC) power is an integral over the distribution with energy $\varepsilon$ of the synchrotron seed photons

$$P_{ic}(\gamma) = -mc^2\frac{d\gamma}{dt} = c\int_0^\infty d\varepsilon \frac{u(\varepsilon)}{\varepsilon}\sigma(\gamma,\varepsilon)\overline{\varepsilon}_{ic}(\gamma,\varepsilon) \tag{28}$$

where

$$u(\varepsilon) = \frac{3}{8\pi}\frac{mc^2}{he}B\tau\int_{\gamma_m}^\infty \frac{dN(\gamma)}{N_e} f_{sy}\left(\frac{\varepsilon}{\varepsilon_{sy}(\gamma)}\right) \tag{29}$$

is the spectral/differential synchrotron energy density, calculated as an integral over the normalized electron distribution of the synchrotron function[2]$f_{sy}(\varepsilon)$,

$$\varepsilon_{sy}(\gamma) = \frac{3he}{16mc^2}B\gamma^2 \tag{30}$$

is the synchrotron characteristic energy, $\tau$ is the source optical-thickness to photon scattering,

$$\sigma(\gamma,\varepsilon) = \frac{3\sigma_e}{8z}\left\{\left[1 - \frac{2(z+1)}{z^2}\right]\ln(2z+1) + \frac{1}{2} + \frac{4}{z} - \frac{1}{2(2z+1)^2}\right\}$$
$$\simeq \sigma_e \begin{cases} (1) \quad 0.45 & z \ll 1 \\ \frac{3}{8z}\left[\ln(2z) + \frac{1}{2}\right] & z \gg 1 \end{cases} \quad z \equiv \frac{\gamma\varepsilon}{mc^2} \tag{31}$$

is the cross-section for scattering a photon of energy $\varepsilon$ by an electron of energy $\gamma$, $\sigma_e$ being the Thomson scattering cross-section and $z$ the photon energy in the electron rest-frame, (the coefficient 0.45 on the first line is for continuity at $z = 1$, the Thomson–Klein-Nishina T-KN transition, and should be used if such scatterings are dominating the iC power), and $\overline{\varepsilon}_{ic}$ is the average energy of the upscattered photon, for which

$$\overline{\varepsilon}_{ic}(\gamma,\varepsilon) \simeq \begin{cases} (4/3)\gamma^2\varepsilon & z \ll 1 \\ \gamma mc^2 & z \gg 1 \end{cases} \tag{32}$$

is a possible approximation (the factor 4/3 should be ignored, to ensure continuity at $z = 1$, if iC cooling is mostly through scatterings at the T-KN transition).

[1] If the magnetic field in the shock's down-stream region is a constant fraction of the equipartition value, then $B^2 \sim u' \sim \Gamma^2 n'$ ($u'$ being the energy density behind the shock, $n'$ the ejecta density ahead of the shock, and $\Gamma$ the shock's Lorentz factor in the frame of the unshocked yet ejecta). The injection rate is $R_i \sim v_{sh} n' \sim cn'$, with $v_{sh}$ being the shock velocity, thus *the condition $R_i \sim B^2$ is satisfied for magnetic fields that are a fixed fraction of the post-shock energy density*

[2] The Band function (Band et al 1993) of low-energy slope $\alpha = -2/3$ and a high-energy slope $\beta \to \infty$ provides a good approximation for the synchrotron function

Equation (28) shows that iC cooling depends on the magnetic field $B$, on the electron injection rate $R_i$ through the optical thickness $\tau \sim N_e = \int R_i dt$, and on the current electron distribution $N(\gamma)$, which itself is determined by the history of the iC cooling power $P_{ic}(\gamma, t)$. This suggests that iC cooling can be accurately calculated only numerically, but some insight about the power-law segments that the electron distribution develops through iC cooling can be obtained analytically.

Substituting the approximations given in equations (31) and (32) in equation (28) leads to

$$P_{ic}(\gamma) = P_t(\gamma) + P_{kn}(\gamma)$$

with

$$\frac{1}{c\sigma_e} P_t(\gamma) \simeq \frac{1}{2}\gamma^2 \int_0^{\varepsilon_k(\gamma)} u(\varepsilon) d\varepsilon \tag{33}$$

and

$$\frac{1}{c\sigma_e} P_{kn}(\gamma) \simeq \frac{3}{8} \int_{\varepsilon_k(\gamma)}^{\infty} u(\varepsilon) \left(\frac{mc^2}{\varepsilon}\right)^2 \left(\ln \frac{2\gamma\varepsilon}{mc^2} + \frac{1}{2}\right) d\varepsilon \tag{34}$$

where

$$\varepsilon_k(\gamma) \equiv \frac{mc^2}{\gamma} > \varepsilon_i \equiv \varepsilon_{sy}(\gamma_i) \tag{35}$$

is the energy of photons that are scattered by electrons of energy $\gamma$ at the T-KN transition. The synchrotron energy density $u(\varepsilon)$ can be calculated as below.

Before the $\gamma_i$ electrons cool significantly, the cooled-injected electron distribution is that of equations (17) and (18), i.e. a broken power-law with a break at $\gamma_c$ and $N(\gamma_i < \gamma < \gamma_c) \sim \gamma^{-q}$ with $q(n < 1) = p + n - 1$ and $q(n > 1) = p$, where $n$ is the to-be-determined exponent of the cooling power $P_{ic}(\gamma) \sim \gamma^n$. The synchrotron energy spectrum corresponding to this electron distribution can be approximated as

$$u(\varepsilon)(t < t_{ci}) = u(\varepsilon_i) \left\{ \begin{array}{ll} (\varepsilon/\varepsilon_i)^{1/3} & (\varepsilon < \varepsilon_i) \\ (\varepsilon/\varepsilon_i)^{-(q-1)/2} & (\varepsilon_i < \varepsilon) \end{array} \right. \tag{36}$$

ignoring for now the steeper power-law segment above the cooling-energy break at $\varepsilon_c \equiv \varepsilon_{sy}(\gamma_c)$.

By equating the synchrotron energy density $\int u(\varepsilon) d\varepsilon$ with the total synchrotron power per unit volume multiplied by the source light-crossing time $P_{sy}\Delta/c = (4/3)\sigma\Delta u_B \int dn_e(\gamma)\gamma^2$, where $u_B = B^2/8\pi$ is magnetic field energy density, it can be shown that

$$\varepsilon_i u(\varepsilon_i) \simeq u_B \gamma_i^2 \tau \tag{37}$$

where $\tau = \sigma_e \Delta n_e$ is the source optical thickness to photon scattering and $n_e$ the electron density.

### 5.1. $\gamma_i$ electrons scattering $\varepsilon_i$ photons in Thomson regime ($z_i < 1$)

First, we consider the case where the $\gamma_i$ scatter their own synchrotron photons of energy $\varepsilon_i$ in the Thomson regime, i.e. $z_i \equiv \gamma_i \varepsilon_i/(mc^2) < 1$. Substitution of the approximate synchrotron spectrum of the second branch of equation (36) in equations (33) and (34) leads to

$$P_t(\gamma) \simeq c\sigma_e \left\{ \frac{1}{3} + \frac{1}{3-q} \left[ \left(\frac{\varepsilon_k}{\varepsilon_i}\right)^{(3-q)/2} - 1 \right] \right\} \gamma^2 \varepsilon_i u(\varepsilon_i) \tag{38}$$

$$P_{kn}(\gamma) \simeq c\sigma_e \left(\frac{mc^2}{\varepsilon_i}\right)^2 \left(\frac{\varepsilon_i}{\varepsilon_k}\right)^{(q+1)/2} u(\varepsilon_i)\varepsilon_i \tag{39}$$

For $q < 3$, the $\varepsilon_k/\varepsilon_i$ term in the Thomson iC power is dominant, then $P_t \simeq P_{kn}$ because

$$\left(\frac{\varepsilon_k}{\varepsilon_i}\right)^{\frac{3-q}{2}} \gamma^2 = \left(\frac{mc^2}{\varepsilon_i}\right)^2 \left(\frac{\varepsilon_i}{\varepsilon_k}\right)^{\frac{q+1}{2}} = \left(\frac{mc^2}{\varepsilon_i}\right)^{\frac{3-q}{2}} \gamma^{\frac{q+1}{2}} \tag{40}$$

(after using from equation 35), and iC cooling is dominated by scatterings at the T-KN transition. If we define the electron energy $\hat{\gamma}_i$ by $\varepsilon_k(\hat{\gamma}_i) = \varepsilon_i$, hence $\hat{\gamma}_i = \gamma_i/z_i > \gamma_i$, (i.e. the synchrotron photons produced by the $\gamma_i$ photons are scattered by the $\hat{\gamma}_i$ electrons at the T-KN transition), it follows that the $\gamma < \hat{\gamma}_i$ electrons cool mostly on $\varepsilon > \varepsilon_i$ photons.

For $q > 3$, the $\varepsilon_k/\varepsilon_i$ term in the Thomson iC power vanishes for $\varepsilon_k \gg \varepsilon_i$, then $P_t \gg P_{kn}$ and iC cooling of $\gamma < \hat{\gamma}_i$ electrons is dominated by scatterings of $\varepsilon_i$ synchrotron photons in the Thomson regime.

The final result is

$$P_{ic}(\gamma < \hat{\gamma}_i, t < t_{ci}) \simeq c\sigma_e u_B \gamma_i^2 \tau \left\{ \begin{array}{ll} \left(\frac{mc^2}{\varepsilon_i}\right)^{\frac{3-q}{2}} \gamma^{\frac{q+1}{2}} & q < 3 \\ \gamma^2 & q > 3 \end{array} \right. \tag{41}$$

#### 5.1.1. $t < t_{ci}$

**Cooled-injected distribution**

Equation (41) shows that, when the iC power is dominated by scatterings in the **Thomson** regime ($q > 3$), $P_{ic}/P_{sy} \simeq Y \equiv \gamma_i^2 \tau$ is the Compton parameter, independent of the electron energy for $\gamma \in (\gamma_i, \hat{\gamma}_i)$, and the cooling-law of equation (3) has

$$K_{ic}(x = 0) \simeq \frac{1}{8\pi} \frac{\sigma}{mc} B^2 \gamma_i^2 \tau \ , \ n = 2 \tag{42}$$

and with same exponent $n = 2$ as for synchrotron emission (equation 25), hence similar conclusions follow: at $t < t_{ci}$, we have $\gamma_i < \gamma_c$ because $n > 1$ (equation 16) and the cooled-injected distribution above $\gamma_i$ is a broken power-law of index $q = p$ for $\gamma_i < \gamma < \gamma_c$ and of index $q = p + 1$ for $\gamma_c < \gamma < \hat{\gamma}_i$. Thus the *iC cooling power is dominated by scatterings in the Thomson regime if the injected electron distribution has an index* $p > 3$.

For an iC power dominated by scatterings at the **T-KN** transition (i.e. for $p < 3$ and $q < 3$), equation (41) shows that the iC power has an exponent $n = (q + 1)/2$; then equation (18) implies that the cooled injected electron distribution has an index $q = p$ below $\gamma_c$ if $n > 1$, which is the case because $n = (q + 1)/2 = (p + 1)/2 > 1$ for $p > 1$. If $\gamma_c < \hat{\gamma}_i$ then, above $\gamma_c$, the cooled-injected electron distribution has an index $q = p + n - 1$ if $n > 1$. We define the electron energy $\tilde{\gamma}_c$ by $\varepsilon_k(\gamma_c) = \varepsilon_{sy}(\tilde{\gamma}_c)$ (i.e. the energy of electrons that radiate synchrotron photons at the T-KN transition for the $\gamma_c$ electrons, where most of the cooling of the $\gamma_c$ electrons occurs), which leads to $\tilde{\gamma}_c = (\gamma_i^3/z_i\gamma_c)^{1/2}$, and note that, for $n > 1$, $\gamma_c$ decreases (equation 16), thus $\tilde{\gamma}_c \sim \gamma_c^{-1/2}$ increases.

For $\tilde{\gamma}_c < \gamma_c$, electrons of energy $\gamma \in (\gamma_c, \hat{\gamma}_i)$ cool mostly by scattering synchrotron photons of energy $\varepsilon \in [\varepsilon_k(\hat{\gamma}_i), \varepsilon_k(\gamma_c)] = [\varepsilon_i, \varepsilon_{sy}(\tilde{\gamma}_c)]$ produced by electrons of energy $\gamma \in (\gamma_i, \tilde{\gamma}_c) \in (\gamma_i, \gamma_c)$, where the cooled electron distribution has the same index as the injected one, $q(\gamma < \gamma_c) = p$, thus the iC power for electrons of energy $\gamma \in (\gamma_c, \hat{\gamma}_i)$ has exponent $n = (q+1)/2 = (p+1)/2$ if $q < 3$ (i.e. if $p < 3$) and the cooled electron distribution in the same energy range will have an index $q = p + n - 1 = (3p-1)/2$ if $n > 1$ (i.e. if $p > 1$).

At later times, when $\gamma_c < \tilde{\gamma}_c$, the above results are valid for electrons energies $\gamma \in (\hat{\gamma}_c, \hat{\gamma}_i)$, where $\hat{\gamma}_c$ is defined by $\varepsilon_k(\hat{\gamma}_c) = \varepsilon_{sy}(\gamma_c)$ (i.e. the $\hat{\gamma}_c$ electrons cool mostly by scattering the synchrotron photons produced by the $\gamma_c$ electrons), thus $\hat{\gamma}_c = \gamma_i^3/z_i\gamma_c^2$ and $\hat{\gamma}_c$ increases. Electrons of energy $\gamma \in (\gamma_c, \hat{\gamma}_c)$ cool mostly by scattering synchrotron photons at their T-KN transition, of energy $\varepsilon \in [\varepsilon_k(\hat{\gamma}_c), \varepsilon_k(\gamma_c)] = [\varepsilon_{sy}(\gamma_c), \varepsilon_{sy}(\tilde{\gamma}_c)] \in [\varepsilon_{sy}(\gamma_c), \varepsilon_{sy}(\hat{\gamma}_c)]$, with the last inclusion folowing from $\tilde{\gamma}_c < \hat{\gamma}_c$, where the cooled electron distribution has an index $q = p = n - 1$ if $n > 1$. Together with $n = (q+1)/2$ for $q < 3$, we obtain $q = 2p - 1$ (hence the working condition $q < 3$ requires $p < 2$) and $n = p$ (thus the working condition $n > 1$ requires $p > 1$). Conversely, for $p > 2$, we expect that $q > 3$, and the iC power has an exponent $n = 2$, leading to a cooled-injected distribution of index $q = p + n - 1 = p + 1 > 3$.

To summarize the above, for $z_i < 1$ and above $\gamma_i$, the electron distribution cooled though iC scatterings develops gradually power-law segments at $\gamma_i - \hat{\gamma}_i$, beginning with a single power-law of index $q = p$, then a new segment with $q = (3p-1)/2$ if $p < 3$ and $q = p + 1$ if $p > 3$ develops at $\gamma_c - \hat{\gamma}_i$ and, after that, another segment with $q = 2p - 1$ if $p < 2$ and $q = p + 1$ if $p > 2$ develops at $\gamma_c - \hat{\gamma}_c$. The lowest and highest energy segments disappear at $t = t_{ci}$ when $\gamma_c = \gamma_i$ and $\hat{\gamma}_c = \hat{\gamma}_i$ thus, leading to a single power-law segment at $\gamma_i - \hat{\gamma}_i$.

The electrons of energy $\gamma > \hat{\gamma}_i$ scatter synchrotron photons of energy $\varepsilon > \varepsilon_i$ in the Klein-Nishina regime, thus their iC cooling is dominated by scattering photons of energy $\varepsilon_k(\gamma) < \varepsilon_i$ at the T-KN transition. The analytical treatment of those electrons' cooling is similar to that in the next section for the $z_i > 1$ case, the result being that $q(\gamma > \hat{\gamma}_i) = p$.

**Cooling-tail**

For $q = p > 3$, the cooling-tail below $\gamma_i$ is a power-law if the injection rate satisfies equation (10): $R_i \sim P_{ic}(\gamma)/\gamma^2 \sim B^2\tau$ (second branch of equation 41). For an electron injection rate $R_i \sim t^y$, the optical thickness is $\tau \sim \int^t R_i dt' \sim t^{y+1}$ for $y > -1$, and the above condition for a power-law cooling-tail is $B^2(y > -1) \sim t^{-1}$. If $y < -1$, then $\tau \simeq const$, and the power-law cooling-tail condition becomes $R_i(y < -1) \sim B^2$, which is naturally satisfied for magnetic fields that are a constant fraction of the shock's energy density.
For $q = p < 3$, the first branch of equation (41) leads to $R_i \sim B^{p-1}\tau$ as the condition for a power-law cooling-tail.

If the cooling-tail is a power-law, then its index $m$ is the exponent $n$ of the cooling power (equation 9), which is given by equation (41), with $q$ being the index of the cooled-injected distribution (discussed above) at the electron energy that produces the synchrotron photons that provide most of the iC cooling of the cooling-tail electrons: $\varepsilon_i$ photons if $q > 3$ and $\varepsilon_k(\gamma) > \varepsilon_i$ photons if $q < 3$. The power-law segments of the cooling-tail following from the power-law segments of the cooled-injected distribution are listed in **Table 1**.

### 5.1.2. $t > t_{ci}$

For $z_i < 1$, the electrons in the cooling-tail at $\gamma < \gamma_i$ cool mostly by scattering synchrotron photons of energy $\varepsilon \geq \varepsilon_i$ ; the energy density of those photons is $u(\varepsilon) \sim \varepsilon^{-(q-1)/2}$ with $q$ being the index of the cooled-injected distribution at $\gamma > \gamma_i$.

For $p > 2$, we have $q = p+1 > 3$ and the synchrotron energy density $\varepsilon u(\varepsilon) \sim \varepsilon^{(3-q)/2}$ peaks at $\varepsilon_i$, thus the cooling-tail electrons cool by scattering the $\varepsilon_i$ photons in the Thomson regime, with the iC cooling power having an exponent $n = 2$ (equation 41). If the cooling-tail is a power-law in electron energy, then its index should be $m = n = 2$.

For $p < 2$, we have $q = 2p - 1 < 3$, and the synchrotron energy density $\varepsilon u(\varepsilon) \sim \varepsilon^{(3-q)/2}$ increases with photon energy, thus the cooling-tail electrons of energy $\gamma$ iC cool mostly by scattering photons of energy $\varepsilon_k(\gamma)$, i.e. photons at the T-KN transition. Then, according to equation (41), the exponent of the cooling power is $n = (q+1)/2 = p$ and, if the cooling-tail is a power-law, then its index will be $m = n = p$.

However, the story does not end here because the cooling-tail developing at $\gamma < \gamma_i$ changes the iC cooling of electrons of energy $\gamma > \hat{\gamma}_i$, which cool mostly through scatterings at the T-KN transition on photons of energy $\varepsilon < \varepsilon_i$, i.e. on photons produced by the cooling-tail. Going further: as a new power-law segment develops at $\gamma > \gamma_i$, it changes the cooling of the lower energy electrons in the cooling-tail that cool mostly through T-KN scatterings of synchrotron photons produced by the electrons in the newly-formed segment at $\gamma > \gamma_i$, leading to the appearance of a new low-energy segment in the cooling-tail. Through this feedback between the development of new segments below $\gamma_i$ (in the cooling-tail) and above $\gamma_i$ (in the cooled-injected distribution), the cooled electron distribution develops more and more breaks and segments.

**Table 1** lists in chronological order the power-law segments expected for the iC-cooled electron distribution, going four more steps beyond what was described above.

It may worth noting that equation (41) is valid only for $t < t_{ci}$, when the cooling-tail has not yet developed and the effective electron distribution can be approximated as a single power-law above $\gamma_i$ of an index $q$ to-be-determined, even though it can develop other power-law segments at $\gamma \gg \gamma_i$. At $t > t_{ci}$, when there is a cooling-tail, the cooled electron distribution can be approximated as a broken power-law, with to-be-determined indices $m$ below $\gamma_i$ and $q$ above $\gamma_i$. That changes the coefficients in equations (33) and (34) and the synchrotron energy density of equation (37), but does not affect the cooling power's $\gamma$-exponents in equation (41), hence it does not change the indices of the cooled electron distribution that were derived above or those listed in Table 1.

Left panels of **Figure 3** show the electron distributions obtained numerically for $z_i < 1$ and $p > 3$, i.e. when the $\gamma \lesssim \gamma_i$ electrons cool mostly through Thomson scatterings on the $\varepsilon_i$ synchrotron photons.

### 5.2. $\gamma_i$ electrons scattering $\varepsilon_i$ photons in KN regime ($z_i > 1$)

For $z_i = \gamma_i \varepsilon_i / mc^2 > 1$, all electrons scatter the synchrotron emission above $\varepsilon_i$ in the Klein-Nishina regime, where the iC power is diminished by the reduced scattering cross-section. In this case, $\varepsilon_k(\gamma) < \varepsilon_i$ and the iC power is dominated by scattering photons below $\varepsilon_i$; using the first branch of equation (36) in equations (33) and (34) leads to

$$P_t(\gamma) \simeq \frac{1}{3} c\sigma_e \left(\frac{\varepsilon_k}{\varepsilon_i}\right)^{4/3} \gamma^2 \varepsilon_i u(\varepsilon_i)$$

$$\frac{P_{kn}(\gamma)}{c\sigma_e \varepsilon_i u(\varepsilon_i)} \simeq \frac{9}{16} \left(\frac{mc^2}{\varepsilon_i}\right)^2 \left[2.2 \left(\frac{\varepsilon_i}{\varepsilon_k}\right)^{2/3} - \ln \left(2 z_i \frac{\gamma}{\gamma_i}\right)\right] \tag{43}$$

For $\varepsilon_k \ll \varepsilon_i$, the logarithmic term can be ignored, the Thomson and KN cooling powers are comparable because

$$\left(\frac{\varepsilon_k}{\varepsilon_i}\right)^{4/3} \gamma^2 = \left(\frac{mc^2}{\varepsilon_i}\right)^2 \left(\frac{\varepsilon_i}{\varepsilon_k}\right)^{2/3} = \left(\frac{mc^2}{\varepsilon_i}\right)^{4/3} \gamma^{2/3} \tag{44}$$

and the electrons cool mostly by scattering $\varepsilon < \varepsilon_i$ photons at the T-KN transition, thus

$$P_{ic}(\gamma, t < t_{ci}) \simeq 1.6 c\sigma_e u_B \gamma_i^2 \tau \left(\frac{mc^2}{\varepsilon_i}\right)^{4/3} \gamma^{2/3} \tag{45}$$

**Cooling-tail**

Equation (45) shows that the iC cooling-law has $n = 2/3$ and $P(\gamma)/\gamma^n \sim (B/\gamma_i)^{2/3}\tau$. Consequently, if the injection rate $R_i \sim t^y$ satisfies equation (10): $R_i \sim B^{2/3}\tau$, which is equivalent to $B(y > -1) \sim t^{-3/2}$ and to $R_i(y < -1) \sim B^{2/3}$ (thus $B(y < -1) \sim t^{-x}$ with $x > 3/2$), then the power-law cooling-tail condition is fullfilled and the *cooling-tail* has an index $m = n = 2/3$. The cooling-tail continues to develop with $m = 2/3$ by scattering at the T-KN transition photons of energy $\varepsilon < \varepsilon_{sy}(\gamma_m)$, whose spectrum is $u(\varepsilon) \sim \varepsilon^{1/3}$, until the lowest energy electrons at $\gamma_m$ scatter their own synchrotron photons at the T-KN transition, i.e. until $\gamma_m \varepsilon_{sy}(\gamma_m) = mc^2$, corresponding to $\gamma_m = \gamma_i / z_i^{1/3}$.

After that, the $m = 2/3$ cooling-tail below $\gamma_i$ shrinks, existing only above an energy $\hat{\gamma}_m$ defined by $\varepsilon_k(\hat{\gamma}_m) = \varepsilon_{sy}(\gamma_m)$, as its electrons continue to cool on synchrotron photons of energy $\varepsilon < \varepsilon_k(\hat{\gamma}_m) = \varepsilon_{sy}(\gamma_m)$. Electrons of energy $\gamma < \hat{\gamma}_m$ cool by scattering photons of energy $\varepsilon > \varepsilon_k(\hat{\gamma}_m) = \varepsilon_{sy}(\gamma_m)$ in the $z = \gamma\varepsilon/mec^2 < 1$ regime presented in previous section. Therefore, their cooling power is similar to that given in equation (41), but with an exponent $n = (m+1)/2$ if $m < 2$. It follows that the cooling-tail has an index $m = n$, thus $m = 1$ (and the $m < 2$ condition is satisfied).

When the increasing electron energy $\hat{\gamma}_m = \gamma_i^3/(z_i \gamma_m^2)$ reaches $\gamma_i$, only the $m = 1$ cooling-tail branch exists, with the $m = 2/3$ segment having disappeared. The $m = 1$ cooling-tail arising from scatterings in the K-N regime has been identified also by Nakar, Ando & Sari (2009) and by Daigne, Bosnjak & Dubus (2011) (who advocated even softer cooling-tails with $m > 1$).

**Cooled-injected distribution**

For a cooling power exponent $n = 2/3 < 1$, equation (16) shows that $\gamma_c < \gamma_i$ at $t < t_{ci}$, and equation (17) indicates that the *cooled-injected* distribution has an index $q = p$. At $t > t_{ci}$, when $\gamma_c > \gamma_i$, same equation (17) shows that the cooled-injected distribution has an index $q = p + n - 1 = p - 1/3$ for electrons of energy $\gamma \in (\gamma_i, \gamma_c)$ and index $q = p$ for $\gamma > \gamma_c$. The cooled-injected distribution continues to be a broken power-law that steepens by 1/3 at the cooling energy $\gamma_c$ as long as the minimal energy electrons $\gamma_m$ in the cooling-tail below $\gamma_i$ reach the energy $\tilde{\gamma}_i$ defined by $\varepsilon_k(\gamma_i) = \varepsilon_{sy}(\tilde{\gamma}_i)$.

That happens when $\tilde{\gamma}_m = \gamma_i$, hence, when the cooling-tail becomes a $m = 1$ power-law, the electrons of energy $\gamma \in (\gamma_i, \tilde{\gamma}_m)$ cool by scattering at the T-KN transition photons of energy $\varepsilon > \varepsilon_k(\tilde{\gamma}_m)\varepsilon_{sy}(\gamma_m)$ produced by the cooling-tail with $m = 1$. The iC power is similar to that given in equation (41) but with exponent $n = (m+1)/2 = 1$. The resulting index of the cooled-injected distribution at $(\gamma_i, \tilde{\gamma}_m)$ should be that in equations (17) and (18), but, for $n = 1$, the cooling energy $\gamma_c$ (equation 16) is not defined. Assuming a continuity of the cooled-injected distribution index across $n = 1$, and noting that $p + n - 1 = p$ for $n = 1$ (i.e. the two branches of equations (17) and (18) are identical), it follows that the distribution index at $(\gamma_i, \tilde{\gamma}_m)$ should be $q = p$.

Above $\tilde{\gamma}_m$, the cooled-injected distribution has the previously derived two branches of indices $q = p - 1/3$ (at $\tilde{\gamma}_m - \gamma_c$) and $q = p$ (at $\gamma > \gamma_c$), resulting from cooling on synchrotron photons of energy $\varepsilon < \varepsilon_{sy}(\gamma_m)$ with an cooling power exponent $n = 2/3$, and a break at $\gamma_c$, which is defined for $n \neq 1$.

Other branches develop below $\gamma_i$ after $\gamma_m$ decreases further and falls below $\hat{\gamma}_i$ defined by $\varepsilon_k(\hat{\gamma}_i) = \varepsilon_{sy}(\gamma_i)$, thus $\hat{\gamma}_i = \gamma_i / z_i < \gamma_i$. As for the $z_i < 1$ case, there is a feedback between branches below and above $\gamma_i$. **Table 2** follows this feedback for two more steps on each side of $\gamma_i$ beyond what was described above.

Right panels of **Figure 3** show the electron distributions obtained numerically for $z_i > 1$, when the $\gamma \gtrsim \gamma_i$ electrons cool mostly through scatterings on the $\varepsilon < \varepsilon_i$ synchrotron photons at the T-KN transition.

## 6. GRB SYNCHROTRON SPECTRUM AND PULSES

Synchrotron emission spectra and light-curves are calculated by integrating the synchrotron function $f_{sy}$ over the electron distribution:

$$F(\varepsilon) = \varepsilon C(\varepsilon) \sim \int d\gamma N(\gamma) \frac{B^2 \gamma^2}{\varepsilon_{sy}(\gamma)} f_{sy}\left(\frac{\varepsilon}{\varepsilon_{sy}(\gamma)}\right)$$

$$\sim B \int d\gamma N(\gamma) f_{sy}\left(\frac{\varepsilon}{\varepsilon_{sy}(\gamma)}\right) \tag{46}$$

where $\varepsilon_{sy}(\gamma)$ is the synchrotron characteristic energy for electrons of energy $\gamma$ (equation 30).

### 6.1. Spectral softening

The synchrotron spectrum is characterized by the peak-energy of the power-per-decade $\varepsilon F_\varepsilon$, the low and high-energy photon spectral slopes $dC/d\varepsilon \sim \varepsilon^{-\alpha}$ below the peak ($\varepsilon < E_p$) and $dC/d\varepsilon \sim \varepsilon^{-\beta}$ above the peak ($\varepsilon > E_p$), and by the Width at Half-Maximum (WHM) of the $\varepsilon F_\varepsilon$ spectrum, which is primarily dependent on the low- and high-energy spectral slopes $\alpha$ and $\beta$, with some contribution from the shape/smoothness of the synchrotron function at its peak, but without any contribution from electron cooling across the injected energy $\gamma_i$: as shown in Figures 1, 2, and 3, the cooled electron distribution has a sharp transition at $\gamma_i$. A small contribution of about 0.10 dex arises from the spread in photon energy due to the differential relativistic boost over the spherically-curved emitting surface (for integrated spectra) or over the ellipsoidal equal arrival-time (for instantaneous spectra).

The progressive broadening and softening of a GRB synchrotron spectrum is shown in **Figure 4**. As the cooling-tail forms at $t > t_{ci}$, the low-energy photon spectral slope $\alpha$ softens toward the expected value. The peak-energy $E_p$ of the $\varepsilon F_\varepsilon$ spectrum also decreases and that evolution becomes faster after electron injection stops (Figures 5 and 6, right upper panels). A spectral softening throughout the pulse and the burst has been observed in bright GRBs, either as a decrease of $E_p$ during the pulse tail (Ford et al 1995) or as a decrease of the hardness ratio (Bhat et al 1994), the ratio of counts at higher energy (above 100 keV) to those at lower energy (below 100 keV).

Below, we derive analytically the evolution of the peak-energy $E_p$ during and after electron injection, for adiabatic and synchrotron cooling, by tracking the evolution of the energy $\gamma_b$ of electrons that radiate at the peak-energy $E_p$ of the synchrotron spectrum.

For **adiabatic** cooling, during the pulse rise ($t < t_i$, with $t_i$ the time when electron injection stops), the peak-energy $E_p$ of the $\varepsilon F_\varepsilon$ spectrum is set by the typical electron energy $\gamma_i$ if $p > 3$, thus $E_p \simeq \varepsilon_{sy}(\gamma_i) \sim \gamma_i^2 B \sim B$ if electrons are injected at a fixed $\gamma_i$. During the pulse decay ($t > t_i$, after electron injection ceases), if the cooling-tail is harder than $m = 3$, then $E_p$ is set by the electron energy $\gamma_b$ resulting from cooling of the $\gamma_i$ electrons from $t_i$ to current time $t$. Integrtaing equation (20) leads to $\gamma_b = \gamma_i (t/t_i)^{-2/3}$, thus $E_p \sim \gamma_b^2 B \sim B t^{-4/3}$. In summary

$$(AD)\ E_p(t) \sim \begin{cases} \gamma_i^2 B & t < t_i \quad (\gamma_b = \gamma_i) \\ \gamma_i^2 B t^{-4/3} & t_i < t \quad (\gamma_b \sim t^{-2/3}) \end{cases} \tag{47}$$

For **synchrotron** cooling, before the $\gamma_i$ electrons cool ($t < t_{ci}$), $E_p$ is set by the cooling energy $\gamma_c$ if $p < 3$ and by $\gamma_i$ if $p > 3$. In the former case, equation (16) with $n = 2$ implies that $\gamma_c \sim t^{x-1} \sim B^{-2} t^{-1}$ for a magnetic field $B \sim t^{-x/2}$, thus $E_p \sim \varepsilon_{sy}(\gamma_c) \sim \gamma_c^2 B \sim B^{-3} t^{-2}$. After the $\gamma_i$ electrons cool ($t > t_{ci}$), but still during electron injection ($t < t_i$), $E_p$ is set by $\gamma_i$ if the cooling-tail is harder than $m = 3$. After electron injection ceases ($t > t_i$), $E_p$ is set by the energy $\gamma_b$ that the injected $\gamma_i$ electrons reach after cooling from injection end-time $t_i$ to current time $t$. Integrating equation (3), one obtains for $\gamma_b$ the same time dependence as for $\gamma_c$: $\gamma_b \sim t^{x-1} \sim (B^2 t)^{-1}$ for $x < 1$, thus $E_p \sim \gamma_b^2 B \sim (B^3 t^2)^{-1}$, and $\gamma_b \simeq const$ for $x > 1$, hence $E_p \sim B$. Summarizing, the pulse shape is given by

$$E_p(t) \sim \begin{cases} B^{-3} t^{-2} & t < t_{ci}\ \&\ p < 3 \quad (\gamma_b = \gamma_c) \\ \gamma_i^2 B & t < t_{ci}\ \&\ p > 3 \quad (\gamma_b = \gamma_i) \\ \gamma_i^2 B & t_{ci} < t < t_i \quad (\gamma_b = \gamma_i) \\ B^{-3} t^{-2} & t_i < t < t_\varepsilon\ (x < 1) \quad (\gamma_b \sim 1/B^2 t) \\ B \sim t^{-x/2} & t_i < t < t_\varepsilon\ (x > 1) \quad (\gamma_b = const) \\ t^{-1} & t_\varepsilon < t\ (lae) \end{cases} \tag{48}$$

and is determined by the evolutions of $\gamma_i$ (assumed to be constant so far) and that of $B$ (assumed to be a power-law).

The last case refers to the evolution of the peak-energy when the received flux is dominated by the emission produced at earlier times by the fluid moving at larger angle, i.e. when the smaller angle-emission switches off at the observing energy either because the magnetic field has turned off or because the observing photon energy is above the intrinsic synchrotron spectrum cut-off created by cooling after electron injection has stopped. That cut-off electron energy can be calculated by integrating equation (3) from $\gamma(t_i) = \infty$ to $\gamma(t_\varepsilon) = \gamma(\varepsilon)$, the energy of electrons that radiate at photon energy $\varepsilon$ equation 30.

The synchrotron emission produced at a comoving frame time $t$ arrives at observer at a time $T(\theta) = t/D$ and is relativistically boosted $F_\varepsilon \sim I_{\varepsilon/D} D^4$ by a Doppler factor $D = [\Gamma(1 - \beta \cos\theta)]^{-1}$ with $\theta$ the angle at which the emitting infinitesimal region on the source's spherical surface moves relative to the direction toward the observer. Thus, most emission arises from a region with $\theta < \Gamma^{-1} \ll 1$, hence $D = 2\Gamma/(1 + \Gamma^2\theta^2)$.

Due to the spherical curvature of the emitting surface, the emission from a region moving at a larger angle $\theta$ arrives later and is less Doppler boosted. If the intrinsic emission from $\theta < \Gamma^{-1}$ falls below the delayed, $\theta > \Gamma^{-1}$ "large-angle" emission, then $E_p = D\varepsilon_{sy}(\gamma_i)$ and $T = t/D$ imply that the peak-energy of the GRB spectrum evolves as $E_p \sim T^{-1}$ (Fenimore et al 1996, Kumar & Panaitescu 2000).

### 6.2. Analytical pulse shape

The evolution of photon flux can be calculated analytically if the synchrotron spectrum is assumed to be a sharp broken power-law at a break-energy $\varepsilon_b \sim \gamma_b^2 B$, of slopes $\alpha$ and $\beta$ below and above $\varepsilon_b$, and with a photon spectral/differential flux at $\varepsilon_b$ of $C_b \sim F_{sy}/\varepsilon_b^2$, where $F_{sy} \sim N_e B^2 \gamma_b^2$ is the synchrotron power and $N_e$ is the number of electron radiating at/above photon energy $\varepsilon_b$. Thus, the approximate photon spectrum is

$$C(\varepsilon) \sim N(> \gamma_b) \begin{cases} \gamma_b^{-2(\alpha+1)} B^{-\alpha} & \varepsilon < \varepsilon_b \\ \gamma_b^{-2(\beta+1)} B^{-\beta} & \varepsilon_b < \varepsilon \end{cases} \tag{49}$$

While electrons are injected ($t < t_i$), the electron break-energy energy is $\gamma_b = \gamma_i = const$; and the number of electrons $N(> \gamma_b)$ is that injected during the last cooling timescale: $N = \int_{max(0,t-t_c)}^{t} R_i(t')dt'$. After the end of electron injection ($t > t_i$), $\gamma_b$ evolves according to equation (3) and $N(> \gamma_b)$ remains constant. For a power-law injection rate $R_i \sim t^y$, we have

$$N(> \gamma_b) \sim \begin{cases} t^{y+1} \sim R_i t & t < t_c \quad y > -1 \\ const & t < t_c \quad y < -1 \\ t^y t_c \sim R_i t_c & t_c < t < t_i \\ const & t_i < t \end{cases} \tag{50}$$

with $t_i$ the time when electron injection ceases.

For **adiabatic** cooling: $t_c \simeq t$ (equation 21), $N \sim R_i t$ for $t < t_i$, $\beta = -(p+1)/2$, and $\gamma_b$ is that of equation (47). Then, equation (49) leads to

$$(AD) \quad C(\varepsilon < \varepsilon_b) \sim B^{-\alpha} \left\{ \begin{array}{ll} R_i t & t < t_i \\ t^{4(\alpha+1)/3} & t_i < t \end{array} \right. \tag{51}$$

$$(AD) \quad C(\varepsilon > \varepsilon_b) \sim B^{(p+1)/2} \left\{ \begin{array}{ll} R_i t & t < t_i \\ t^{-2(p-1)/3} & t_i < t \end{array} \right. \tag{52}$$

with $\alpha$ given in equation (24).

For **synchrotron** cooling, the evolution of $\gamma_b$ is given in equation (48), the photon spectral slope just above the break $\varepsilon_b$ is $\beta = -(p+1)/2$ at $t < t_{ci}$ (uncooled electrons) and $\beta = -(p+2)/2$ at $t > t_{ci}$ (cooled electrons) (see Figure 2, lower left panel), and the photon rate is

$$(SY) \quad C(\varepsilon < \varepsilon_b) \sim \left\{ \begin{array}{ll} R_i t B^{3/2} & t < t_{ci} \\ R_i B^{-1/2} & t_{ci} < t < t_i \\ B^{-1/2} t^{-1} & t_i < t < t_\varepsilon \ (x < 1) \\ B^{3/2} & t_i < t < t_\varepsilon \ (x > 1) \\ t^{\alpha-1} & t_\varepsilon < t \ (lae) \end{array} \right. \tag{53}$$

where $B \sim t^{-x/2}$

$$(SY) \quad C(\varepsilon > \varepsilon_b) \sim \left\{ \begin{array}{ll} R_i t B^{(p+1)/2} & t < t_{ci} \\ R_i B^{(p-2)/2} & t_{ci} < t < t_i \\ B^{(p-2)/2} t^{-1} & t_i < t < t_\varepsilon \ (x < 1) \\ B^{(p+2)/2} & t_i < t < t_\varepsilon \ (x > 1) \\ t^{-(p+4)/2} & t_\varepsilon < t \ (lae) \end{array} \right. \tag{54}$$

where $t_\varepsilon$ is the epoch *after* electron injection ends when the received flux is dominated by the large-angle emission (see after equation 48). The above lae decays follow from the Doppler boost of the intronsic emission $C_\varepsilon \sim I_{\varepsilon/D} D^3$, for a comoving frame spectrum $I_\varepsilon \sim \varepsilon^{\alpha(\beta)-1}$ (Kumar & Panaitescu 2000).

Equations (51)–(54) give the pulse shape, with a rise at $t < t_i$ and a fall at $t > t_i$, at energies below and above the spectrum peak $E_p = \Gamma \varepsilon_b$, and for any injection rate $R_i(t)$ and magnetic field $B(t)$ histories For power-law evolving $R_i$ and $B$, the 0.3-0.5 dex energy-band pulses shown in Figures 5 and 6 display a shape roughly compatible to these analytical results, but their accuracy is significantly reduced by the assumption of a sharply broken power-law spectrum at $\varepsilon_b$. As Figure 4 shows, the transition between the low-energy and high-energy power-laws is smooth and the observing energy channels are often close to the the peak-energy $E_p$.

### 6.3. Numerical synchrotron pulses

According to Nemiroff et al (1994), Norris et al (1995), and Lee, Bloom & Petrosian (2000) most GRB pulses are time-asymmetric, with a rise-time $T_r$ shorter than the fall-time $T_f$, peak earlier and last shorter at higher energy, and have a time-asymmetry ratio $T_r/T_f$ that is independent of photon energy.

The pulse rise could be due to the accumulation of electrons that radiate in the observing band, while the pulse-tail could occur after electron injection ceases, as electrons cool below the observing window. Alternatively, the evolution of the magnetic field may determine most of the pulse shape. That pulses peak earlier and are shorter at higher energy could be due to the continuous electron cooling and to the geometrical curvature of the emitting spherical surface, which makes photons arriving later at observer have a lower energy. Additionaly, these two pulse spectral-temporal correlations could be due to a decreasing magnetic field.

The spectral softening, shape, duration, and dependence on photon energy of synchrotron pulses are shown in **Figure 5** for adiabatic electron cooling, and in **Figure 6**, for synchrotron cooling. Unless the magnetic field increases fast, the peak-energy $E_p$ of the synchrotron spectrum is decreasing throughout the pulse (Figures 5 and 6, upper right panels), which leads to peaks occurring earlier and pulses lasting shorter at higher photon energy, as is observed in GRBs. For either adiabatic or synchrotron electron cooling, the resulting pulse duration decrease with photon energy (Figures 5 and 6, lower right panels) is consistent quantitatively with observations: WHM$\sim E^{-0.4}$ (Fenimore et al 1995, Norris et al 1996).

For adiabatic-dominated cooling, only a constant or slowly decreasing injection rate $R_i$ or magnetic field $B$ yield the observed pulse time-asymmetry (Figure 5, left upper panel). Pulses become more time-symmetric at higher energies (Figure 5, left lower panel), which is inconsistent with most observations, albeit such a trend is reported by Norris et al (1995) when comparing the pulse time-asymmetry in the two lowest energy channels (25–100 keV) and the two highest channels (100–1000 keV), for pulses separated in duration and brightness groups.

In the case of synchrotron cooling, the pulse obtained for a constant $R_i$ and $B$ is too time-symmetric, thus the observed pulse time-asymmetry constrains power-law evolving $R_i \sim t^y$ and $B \sim t^{-x/2}$ to be around those shown in Figure 6 (left upper panel). Pulses display the observed energy-independent time-asymmetry (Figure 6, left lower panel).

## 7. CONCLUSIONS

The evolution of an electron distribution is described by the conservation equation (1) and the electron cooling by the cooling-law of equation (3). In this work, solutions to these equations were derived analytically and calculated numerically under the assumption that the injected electron distribution has a fixed minimal/typical electron energy $\gamma_i$ and that the injected distribution is a power-law of fixed exponent $p$.

The index $m$ of the power-law *cooling-tail* (below the typical injected energy $\gamma_i$) depends on the electron injection rate $R_i$ and magnetic field $B$ histories, and was calculated analytically and confirmed numerically under the assumption the $R_i(t)$ and $B(t)$ are power-laws. The cooling-tail index $m$ determines the slope $\alpha$ of the low-energy synchrotron spectrum: $dC/d\varepsilon \sim \varepsilon^\alpha$. Consequently, it is worth investigating for each electron cooling mechanism (adiabatic, synchrotron, inverse-Compton), the conditions that lead to a low-energy photon slope $\alpha$ compatible with those measured by Fermi-GBM, Swift-BAT, or CGRO-BATSE. For the last, Preece et al (2000) have reported that $\alpha$ has a normal distribution peaking extending from -2 to 0 and peaking at -1. Further constraints on power-laws $R_i(t)$ and $B(t)$ are imposed by the observed time-asymmetry of GRB pulses, displaying a shorter rise than the fall timescale.

Other salient properties of GRB pulses (peaks occurring earlier and lasting shorter at higher energy) and the progressive softening of the GRB pulse spectrum (manifested either as a softening of the low-energy photon slope $\alpha$ or as a decrease of the peak energy $E_p$) were shown to arise naturally from electron cooling, with some contribution from the spherical curvature of the emitting surface and, possibly, from a decreasing magnetic field.

### 7.1. GRB low-energy spectrum

For any cooling process, the cooling-tail that develops below the minimal injected electron energy $\gamma_i$, is a pure power-law only for particular electron injection rates $R_i(t)$, of increasing complexity as one goes from adiabatic losses, with the important feature that the *cooling-tail power-law index* $m$ *is the same as the exponent* $n$ at which the electron energy appears in the cooling-law/power (equations 3 and 9), if $n \neq 1$.

For **adiabatic** cooling, where $n = 1$, the index $m$ cannot be determined by solving the conservation equation (1). Continuity of solutions for that equation suggests that $m = 1$ for $n = 1$, but other indices $m$ are also possible for adiabatic cooling. Continuity of the electron distribution at $\gamma_i$ leads to the simplest constraint on $R_i$ for which the cooling-tail is a power-law: a power-law $R_i(t) \sim t^y$, and the index $m = -(3y+1)/2$.

For **synchrotron** cooling, where $n = 2$, then a power-law cooling-tail of index $m = 2$ and a corresponding slope $\alpha = -(m+1)/2 = -3/2$ require $R_i \sim B^2$ (more complex than for adiabatic cooling). That condition is satisfied if the magnetic field is a constant fraction of the internal energy density of the shocked fluid, i.e. $B^2 \sim n'$ with $n'$ being the ejecta density ahead of the shock that energizes it, given that, for a shock moving at constant speed (near $c$), the injection rate $R_i \sim n'$. Compared to histories satisfying the $R_i \sim B^2$ condition, increasing (decreasing) injection rates or decreasing (increasing) magnetic fields lead to harder (softer) non–power-law cooling-tails (Figure 2). leading to GRB low-energy spectral slopes that are harder or softer than $\alpha = -3/2$.

For **inverse-Compton** cooling, power-law cooling-tails result under more restrictive conditions on the injection rate, of the form $R_i \propto B^a \gamma_m^b \tau$, where $\tau \sim \int R_i dt$ is the electron optical thickness to photon scattering, and $\gamma_m$ the lowest energy of the electron distribution (thus $\gamma_m \simeq \gamma_i$ at $t < t_{ci}$, before the injected electron of lowest energy $\gamma_i$ cool significantly). Then, there are four expected cooling-tail indices: $m = 2/3, 1, 2, p$ (Tables 1 and 2), with corresponding low-energy photon indices $\alpha = -5/6, -1$ for any injected index $p$, $\alpha = -(p+1)/2 > -3/2$ if $p < 2$, and $\alpha = -3/2$ if $p > 2$. The first case $m = 2/3$ and $\alpha = -5/6$ is short-lived/transient, the remaining three are persistent. As for synchrotron, different low-energy indices and non–power-law photon spectra result if the injection rate has a different evolution.

To all these results, we add the hardest photon slope $\alpha = -2/3$ (the synchrotron limit) resulting for an uncooled electron distribution or for a cooling-tail harder than $m = 1/3$ (e.g. Figure 2).

### 7.2. GRB high-energy spectrum

With the above two assumptions for the distribution of injected electrons (fixed lowest energy $\gamma_i$ and fixed index $p$), the *cooled-injected* electron distribution above $\gamma_i$, given in equations (17) and (18) for radiative cooling, should be a broken power-law that steepens at the cooling electron energy (equation 16), satisfying the following: $i)$ for electrons with a cooling timescale longer than the time since beginning of injection, the cooled distribution is that injected (of index $p$) $ii)$ for electrons with a cooling timescale shorter than the system's age, cooling (of any origin) yields a power-law distribution of index $q = p + n - 1$, where $n$ is the exponent of the cooling power (equation 3).

Thus:
$i)$ if $n < 1$ (as for a particular case of inverse-Compton cooling dominated by scatterings at the Thomson–Klein-Nishina transition), the cooling timescale $t_c(\gamma) \sim \gamma^{1-n}$ increases with the electron energy $\gamma$ and the cooled electron distribution is flatter than the injected one
$ii)$ if $n = 1$ (as for adiabatic cooling or for another particular case of iC cooling at the T-KN transition), cooling shifts the injected distribution without changing its index
$iii)$ if $n > 1$ (as for synchrotron cooling and iC cooling dominated by scattering in the Thomson regime), the cooling timescale $t_c$ decreases with $\gamma$ and the cooled distribution is steeper than that injected.

The census of indices for the power-law cooled-injected distribution is :
$i)$ adiabatic cooling yields only one index $q = p$
$ii)$ synchrotron cooling yields $q = p$ for pulses that last shorter than the cooling timescale $t_{ci}$ of the $\gamma_i$ electrons and $q = p + 1$ if the pulse spectrum is integrated over more than $t_{ci}$
$iii)$ as for synchrotron, inverse-Compton yields $q = p$ for pulses shorter than the cooling timescale, and four other values $q = p - 1/3, p$ for any $p$, $q = 2p - 1$ if $p < 2$, and $q = p + 1$ if $p > 2$ (Tables 1 and 2), with the first value being transient and the remaining three, persistent.

Thus, the diversity measured for the GRB high-energy photon slopes, which, for BATSE bursts (Preece et al 2000), stretches from -3.5 to -1.5, with a peak at $\beta = -2.25$, points to a range of values for the index $p$ of the injected electron distribution, a conclusion previously reached by Shen, Kumar & Robinson (2006) from GRB spectra and by Starling et al (2008) and Curran et al (2010) from afterglow observations.

### 7.3. Spectral softening and pulse shape

The continuous electron cooling through the observing window always leads to a softening of the low-energy spectral slope $\alpha$ (Figure 4). When electrons are injected, i.e. during the pulse rise, a decrease of the peak-energy $E_p$ of the $\varepsilon F_\varepsilon$ spectrum (Figures 5 and 6, right upper panels) requires a decreasing $B$ (equations 47 and 48). For either cooling mechanism, electron cooling yields a decreasing $E_p$ after electron injection stops, i.e. during the pulse decay. Such a spectral softening is qualitatively consistent with the gradual decrease of the high-to-low energy photon-count ratio that is reported by Bhat et al (1994) and quantitatively consistent with the decrease of $E_p$ by a factor few/several during the pulse tail reported by Ford et al (1995).

Time-asymmetric pulses, displaying a rise faster than the decay, may arise from adiabatic electron cooling, with a significant contribution toward the observed pulse time-asymmetry coming from the spherical curvature of the emitting surface (Figures 5 and 6, left upper panels). Because the pulse shape is determined by the evolution of the electron injection rate $R_i$ and of the magnetic field $B$ (equations 51–54), the observed pulse time-asymmetry can be used as a tool to identify simple $R_i(t)$ and $B(t)$ histories that could be at work in GRBs. For adiabatic cooling, we find that pulses become more time-symmetric at higher photon energy, which may be incompatible with GRB observations, while for synchrotron cooling, the pulse time-symmetry is energy-independent, in accord with observations (Norris et al 1996, Lee et al 2000).

The continuous electron cooling also makes pulses peak earlier and last shorter at higher energies (Figures 5 and 6, lower panels). A significant contribution to these temporal-spectral correlations arises from the curvature of the emitting spherical surface (Figure 6, lower left panel), which reduces the energy of photons that arrive later at observer. The resulting peak-time shift and pulse-duration decrease with energy are quantitatively in accord with GRB observations (Fenimore et al 1995, Norris et al 1996).

### 7.4. **Adiabatic vs radiative cooling**

The above suggests ways to distinguish adiabatic from radiative cooling in GRB spectra and pulses. One is that, owing to that a cooling break exists only for radiative-dominated cooling, a slower decrease of the peak energy $E_p$ during the pulse rise is more likely associated with an adiabatic electron cooling, as illustrated in Figures 5 and 6. The same figures illustrate that the fractional shift of the pulse peak-time (i.e. relative to the pulse duration) is smaller for adiabatic cooling than for synchrotron cooling. That the pulses become more time-symmetric with increasing energy when electron cooling is dominated by adiabatic losses, while pulse appear to have an energy-independent time-asymmetry when synchrotron cooling is dominant, may provide another way to identify the cooling process.

Because the adiabatic cooling timescale is proportional to the source radius, adiabatic cooling is more likely to be dominant for smaller source radii. Together with the source radius being one factor that sets the pulse duration, that implies that adiabatic cooling is more likely to be dominant in shorter pulses. Thus, the signatures of adiabatic cooling (smaller fractional peak-time shifts, pulses becoming more time-symmetric at higher energies) may appear more often in short pulses or in short bursts.

TABLE 1

Indices $m$ and $q$ (equation 4) of the various power-law segments of the electron distribution that develop from a single power-law injected distribution of index $p$ through **inverse-Compton** cooling, for the case where the $\gamma_i$ electrons scatter their own synchrotron emission in the **Thomson** regime ($z_i < 1$). At $t < t_{ci}$, the cooling-tail below $\gamma_i$ (indices listed in square brackets) is not yet fully developed.

| decreasing $\gamma < \gamma_i$ toward left | | | increasing $\gamma > \gamma_i$ toward right | | | | |
|---|---|---|---|---|---|---|---|
| | | | $t < t_{ci}$ | | | | |
| | | [(p+1)/2 $p < 3$]<br>[2 $3 < p$] | p | p | | | |
| | | [(3p+1)/4 $p < 3$]<br>[2 $3 < p$] | p | (3p-1)/2 $p < 3$<br>p+1 $3 < p$ | p | | |
| | | [p $p < 2$]<br>[2 $2 < p$] | p | 2p-1 $p < 2$<br>p+1 $2 < p$ | (3p-1)/2 $p < 3$<br>p+1 $3 < p$ | p | |
| segments develop progressively toward left | | | $t_{ci} \leq t$ | segments develop progressively toward right | | | |
| (11p+5)/16 $p < 2$<br>(10p+7)/16 $p \in (2, 2.5)$<br>2 $2.5 < p$ | (3p+1)/4 $p < 2$<br>(2p+3)/4 $p \in (2, 2.5)$<br>2 $2.5 < p$ | p $p < 2$<br>2 $2 < p$ | 2p-1 $p < 2$<br>p+1 $2 < p$ | (3p-1)/2 $p < 2$<br>p+1/2 $2 < p$ | (11p-3)/8 $p < 2$<br>(10p-1)/8 $p \in (2, 2.5)$<br>p+1/2 $2.5 < p$ | (43p-11)/32 $p < 2$<br>(42p-9)/32 $p \in (2, 2.5)$<br>p+1/2 $2.5 < p$ | p |

TABLE 2

Indices for inverse-Compton cooling when $\gamma_i$ electrons scatter their own synchrotron emission in the **Klein-Nishina** regime ($z_i > 1$)

| decreasing $\gamma < \gamma_i$ toward left | | | increasing $\gamma > \gamma_i$ toward right | | | | |
|---|---|---|---|---|---|---|---|
| | | | $t < t_{ci}$ | | | | |
| | | [2/3] | p | | | | |
| | | | $t_{ci} < t$ | | | | |
| | 1 | 2/3 | p-1/3 | p | | | |
| | | 1 | p | p-1/3 | p | | |
| segments develop progressively toward left | | | | segments develop progressively toward right | | | |
| (5p+3)/8 $p < 2.6$<br>2 $2.6 < p$ | (p+1)/2 $p < 3$<br>2 $3 < p$ | 1 | p | (5p-1)/4 $p < 3$<br>p+1/2 $3 < p$ | (21p-5)/6 $p < 2.6$<br>p+1/2 $2.6 < p$ | p-1/3 | p |

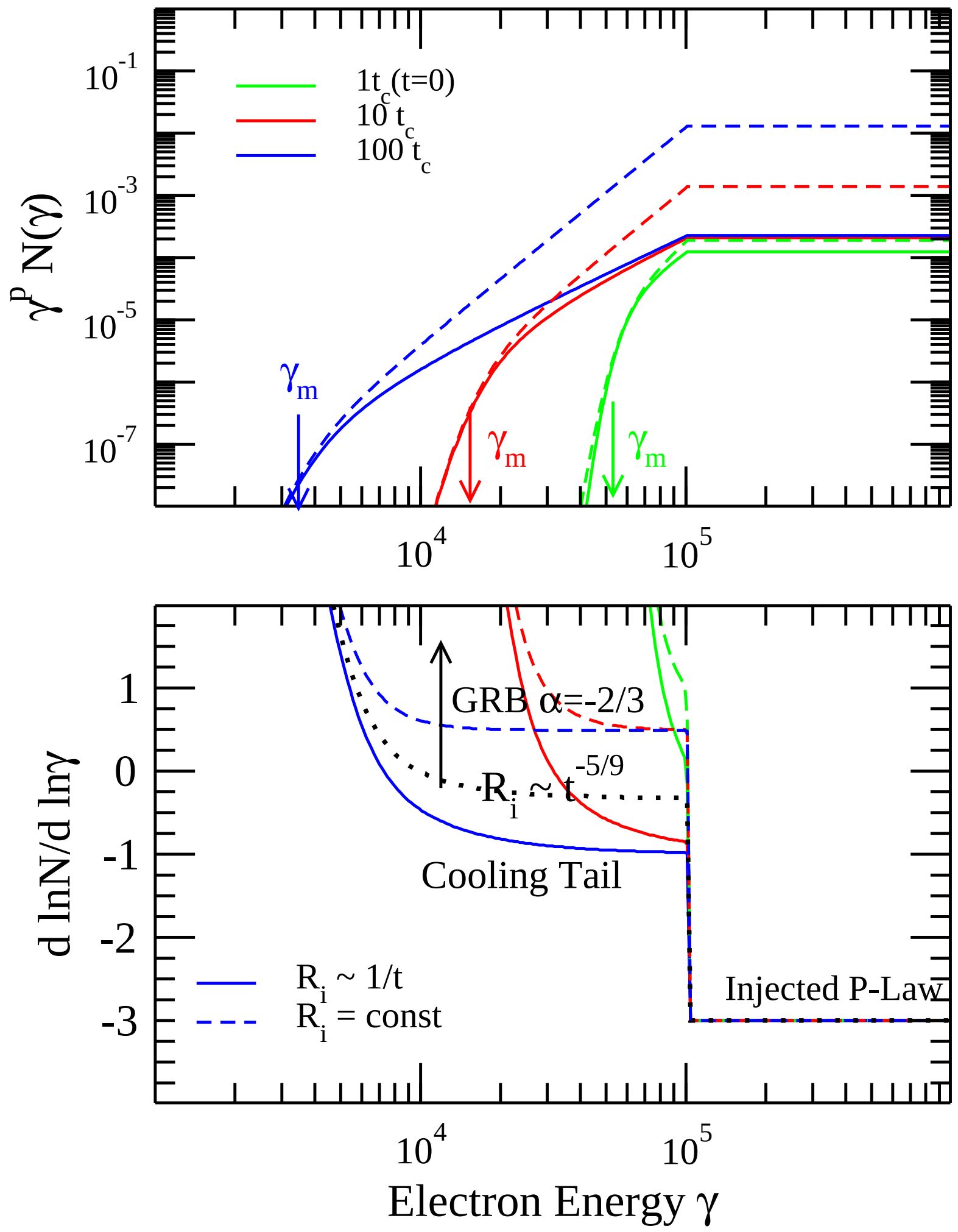

FIG. 1.— Electron distribution (upper panel) multiplied by $\gamma^p$ and its logarithmic slope (lower panel) resulting from cooling **adiabatically** a power-law injected distribution $N(\gamma) \sim \gamma^{-p}$ with $p = 3$ above $\gamma_i = 10^5$, and at three epochs measured in units of the initial cooling timescale $t_c(t = 0) = 1.5t_o$, with $t_o$ the time when electron injection began. Dashed lines are for a constant electron injection rate $R_i$, solid lines are for $R_i \sim 1/t$. The aim here is to verifying the analytical results (equation 23) obtained for an injection rate $R_i \sim t^y$ : $i)$ normalization of the cooled electron distribution is $a(t) \sim A(t) \sim t^{y+1}$, $ii)$ power-law cooling-tail (below $\gamma_i$) $N(\gamma) \sim \gamma^{(3y+1)/2}$, $iii)$ power-law cooled-injected distribution (above $\gamma_i$) $N(\gamma) \sim \gamma^{-p}$. The power-law cooling-tail extends to nearly the minimal electron energy $\gamma_m(t) = \gamma_i(1+t/t_o)^{-2/3}$, indicated in the upper panel with an arrow. Note that adiabatic cooling does not spread the break at $\gamma_i$ of the injected electron distribution. For injection rates with $y < -5/9$ (black, dotted line in lower panel), the cooling-tail is harder than $N(\gamma) \sim \gamma^{-1/3}$ and its synchrotron spectrum is the $F_\varepsilon \sim \varepsilon^{1/3}$ hardest limit for an optically thin (to synchrotron self-absorption) source, i.e. the GRB low-energy photon spectrum has a slope $\alpha = -2/3$ independent of $y$.
Near the lowest cooled electron energy $\gamma_m$, the diffusion arising from the numerical integration of a discontinuous (at $\gamma_i$) injected electron distribution leads to a diverging slope.

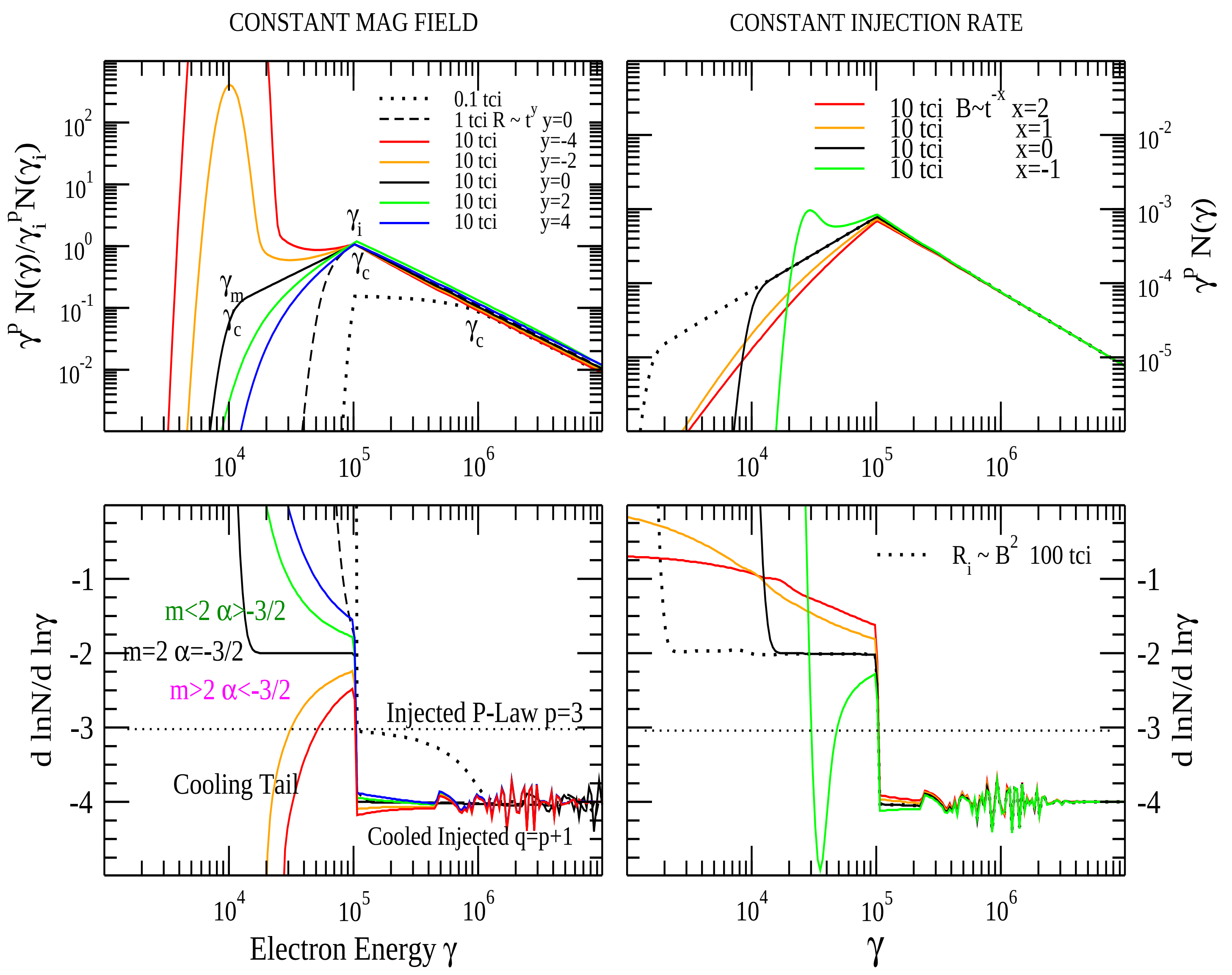

FIG. 2.— **Left**: Electron distribution and its slope resulting from **synchrotron** cooling of an injected electron distribution that is a power-law of index $p = 4$ above $\gamma_i = 10^5$, for a *constant* magnetic field $B = 10$ G and an evolving injection rate $R_i(t) \sim t^y$. Before the $\gamma_i$ electrons cool (e.g. $t = 0.1t_{ci}$), the cooled-injected distribution has a break at $\gamma_c(t) = \gamma_i(t_{ci}/t)$, where the power-law distribution steepens from $N(\gamma < \gamma_c) \sim \gamma^{-p}$ to $N(\gamma > \gamma_c) \sim \gamma^{-(p+1)}$ (dotted curves). At $t > t_{ci}$, the electron distribution develops a tail down to $\gamma_m = \gamma_c = \gamma_i(t_{ci}/t)$. For a constant $R_i$, that tail is a power-law: $N(\gamma) \sim \gamma^{-m}$ with $m = 2$ (black solid curve). An increasing $R_i$ ($y > 0$) yields a harder cooling-tail $N(\gamma < \gamma_i) \sim \gamma^{-m}$ with $m < 2$ (and a harder GRB low-energy spectrum), while a decreasing $R_i$ ($y < 0$) leads to a softer cooling-tail with $m > 2$ (and a softer low-energy GRB spectrum). **Right**: Electron distribution for a *constant* $R_i$ and an evolving $B = 10\,(t/t_{ci})^{-x}$ G with $t_{ci} \equiv t_{ci}(B = 10\,G)$. A decreasing $B$ ($x > 0$) yields a harder cooling-tail, while an increasing $B$ ($x < 0$) leads to a softer tail. If $R_i \sim B^2$ ($y = -2x$), then the effects of evolving $R_i$ and $B$ cancel out and the cooling-tail is a power-law of index $m = 2$ (dotted curve).

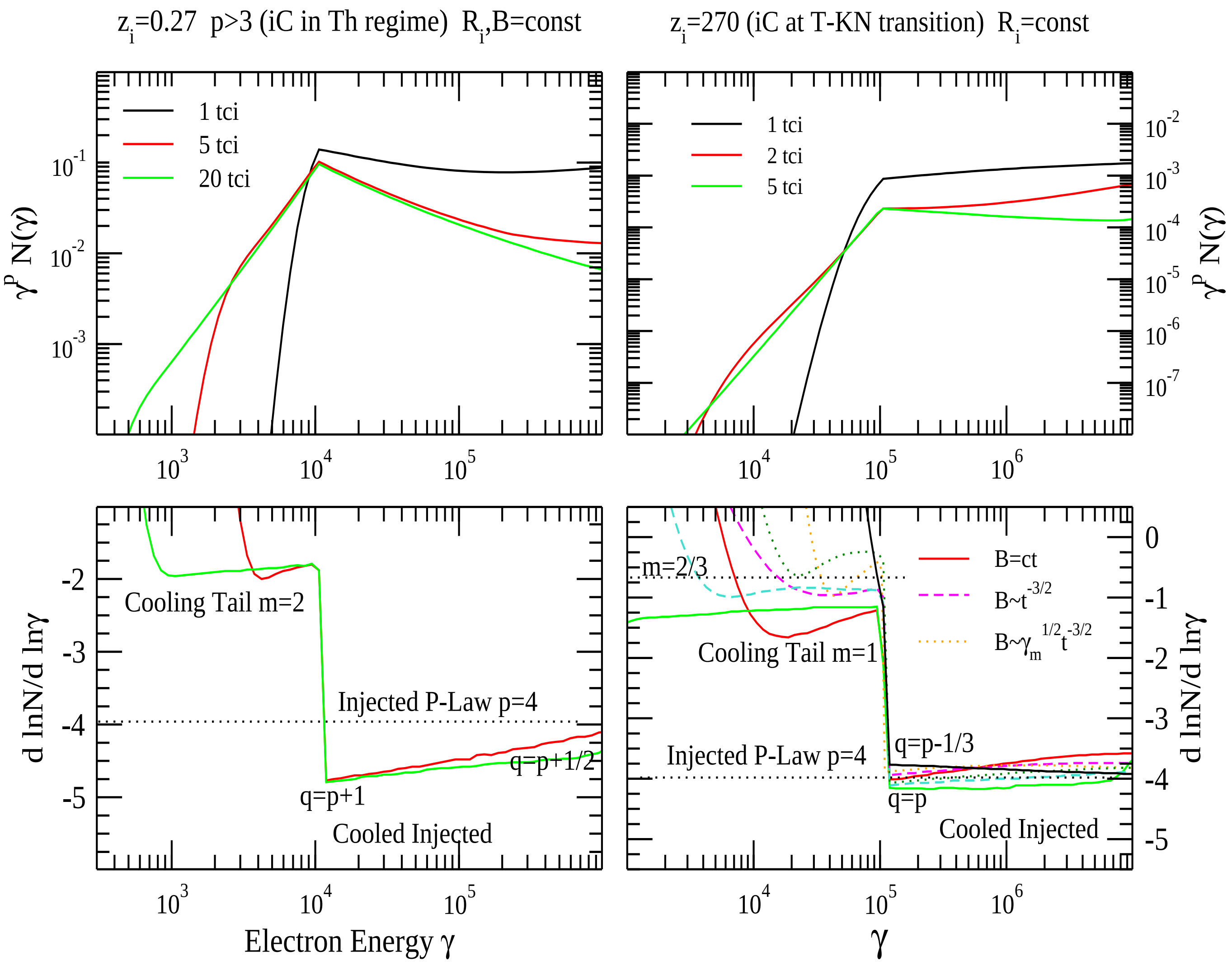

FIG. 3.— **LEFT**: Electron distribution and its slope resulting from **inverse-Compton** cooling of an injected power-law distribution with index $p = 4$ above $\gamma_i = 10^4$, for a magnetic field $B = 10$ G, The electron injection rate $R_i$ is const and yields an electron optical thickness $\tau = 10^{-8}$ and a Compton parameter $Y(\gamma_i) = \gamma_i^2 \tau = 1$ at $t = t_{ci}$. The above $\gamma_i$ and $B$ imply $z_i \equiv \gamma_i \varepsilon_i / mc^2 = 0.27 < 1$, thus the $\gamma_i$ electrons scatter their emission in the **Thomson** regime. The iC power given in equation (41) for $q > 3$ implies that the *cooling-tail* should be a power-law of index $m = 2$, if the condition for a power-law cooling-tail (equation 10) is staisfied. It can be shown that, after $t_{ci}$, the iC power is $P_{ic}(\gamma < \gamma_i) \sim \gamma_m \tau B^2 \gamma^2$, where $\gamma_m$ is the lowest energy in the cooling-tail. By integrating the cooling of $\gamma_m$ electrons: $P_{ic}(\gamma_m) \sim B^2 \tau \gamma_m^3$ with $B = const$ and $\tau \sim t$ (for constant $R_i$), one obtains $\gamma_m \sim t^{-1}$ and the power-law cooling-tail condition $R_i \sim \gamma_m B^{2/3} \tau$ is satisfied. After $t_{ci}$, the *cooled-injected* distribution displays an index $q = p + 1$ above $\gamma_i$, changing to $q = p + 1/2$ at higher energies (Table 1). **RIGHT**: **inverse-Compton** cooling for $\gamma_i = 10^5$, $B = 10$ G (solid lines), constant $R_i$. For these parameters, $z_i \equiv \gamma_i \varepsilon_i / mc^2 = 270 \gg 1$, thus the $\gamma_i$ electrons cool mostly by scattering synchrotron photons of energy $\varepsilon < \varepsilon_i$ at the **Thomson–Klein-Nishina** transition. At $t < t_{ci}$, the iC power of equation (45) iplies a *cooling-tail* of index $m = 2/3$. It can be shown that, after $t_{ci}$, $P_{ic}(\gamma < \gamma_i) \sim \gamma_m^{-1/3} \tau B^{2/3} \gamma^{2/3}$, the condition for a power-law cooling-tail is $R_i \sim \gamma_m^{-1/3} B^{2/3} \tau$, leading to $B \sim \gamma_m^{1/2} t^{-3/2}$, after using $\tau \sim R_i t$. Constant $R_i$ and $B$ do not satisfy this condition and the $m = 2/3$ cooling-tail is not obtained numerically; instead, the cooling-tail displays an index $m \gtrsim 1$. If the condition $B \sim t^{-3/2}$ for a power-law cooling-tail at $t \lesssim t_{ci}$ is enforced (dashed lines), then the cooling-tail displays the $m = 1$ index expected at later times (Table 2), because that $B(t)$ is close to the power-law cooling-tail requirement $B \sim t^{-1}$ corresponding to an iC cooling power $P_{ic}(\gamma < \gamma_i) \sim \tau B \gamma$ at $t > t_{ci}$. Only when the contrived condition $B \sim \gamma_m^{1/2} t^{-3/2}$ is imposed, the numerical cooling-tail displays the expected index $m = 2/3$ (dotted curves). Note: for the above cooling power of exponent $n < 1$, equation (16) shows that $\gamma_c > \gamma_i$ at $t > t_{ci}$. After $t_{ci}$, the *cooled-injected* distribution evolves as expected (Table 2), from $q(\gamma_i < \gamma < \gamma_c) = p - 1/3$ and $q(\gamma_c < \gamma) = p$ to $q(\gamma_i < \gamma < \gamma_c) = p$ and $q(\gamma_c < \gamma) = p - 1/3$.

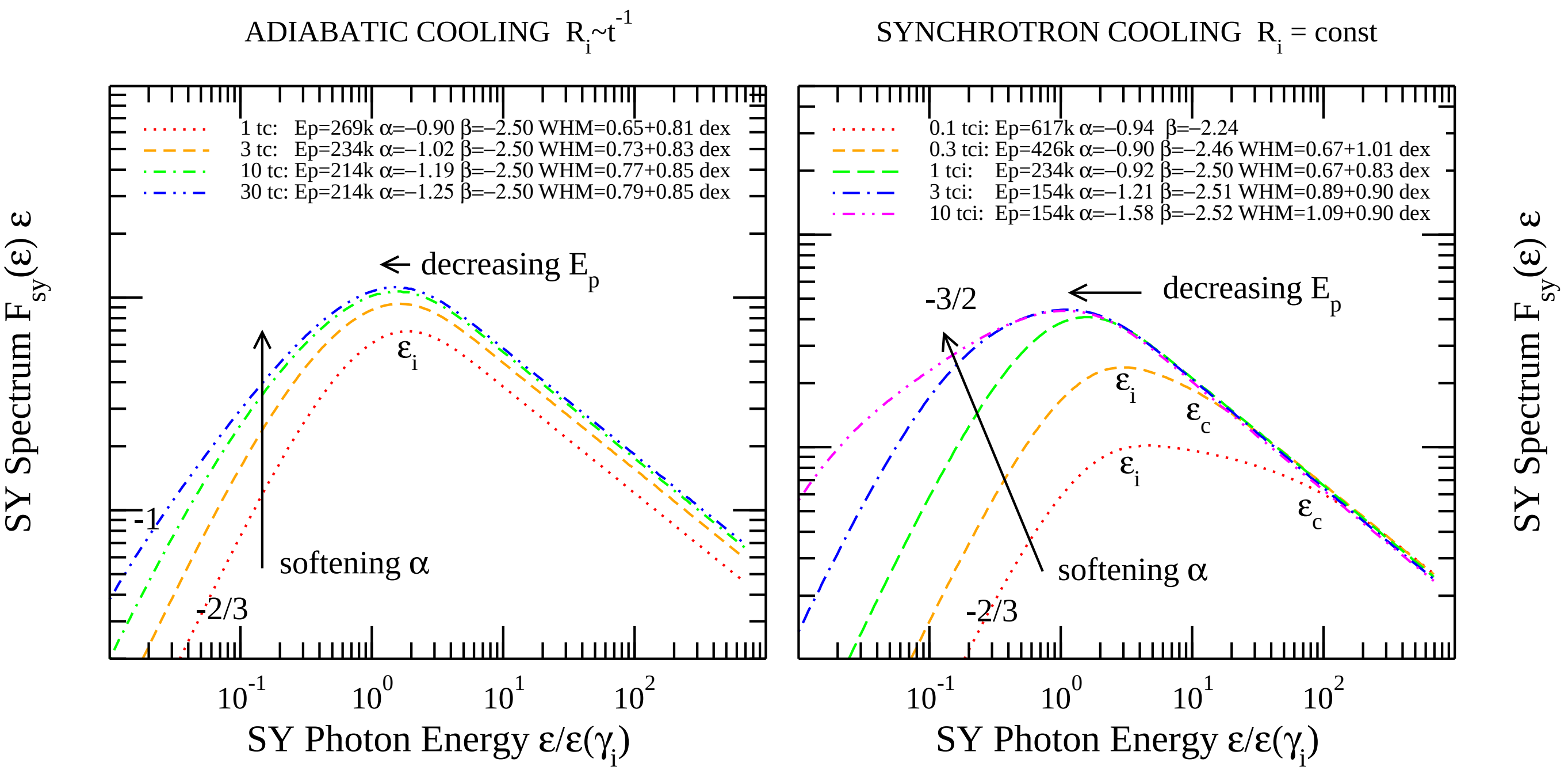

FIG. 4.— Evolution of **synchrotron instantaneous** spectrum for a constant magnetic field ($B = 10$ G). Other parameters: fixed electron injection energy $\gamma_i = 10^5$, source Lorentz factor $\Gamma = 100$. *Definitions*: $E_p$ is the peak of the $\varepsilon F_\varepsilon$ synchrotron spectrum, $\alpha$ and $\beta$ are the photon spectral slopes $-d\ln C/d\ln\varepsilon$ at $0.1\,E_p$ and $10\,E_p$, respectively (these values span the usual GRB observing window 25 keV–1 MeV; however, the synchrotron spectrum may be softer at $0.1E_p$ than the asymptotic value at $E \ll E_p$), WHM is the logarithmic width of the $\varepsilon F_\varepsilon$ spectrum at half of its maximal value (with the width of the rising and falling parts specified separately). **LEFT**: **adiabatic cooling** of a power-law electron distribution of index $p = 4$ with an injection rate $R_i \sim t^{-1}$, which leads to a cooling-tail $N(\gamma < \gamma_i) \sim \gamma^{-1}$ (equation 23, Figure 1) and a corresponding asymptotic low-energy slope $\alpha = -1$. Above $\gamma_i$, the effective electron distribution is that injected (of index $p$). $R_i \sim t^{-1}$ leads to a constant normalization of the electron distribution after $t_c$ (equation 23, see Figure 1), hence the nearly constant spectrum peak flux. Epochs are in units of initial cooling timescale $t_c = 1.5t_o$ (equation 21). **RIGHT**: **synchrotron cooling**, constant $R_i$, and $p = 3$. Epochs are in units of $t_{ci}$, the cooling timescale of $\gamma_i$ electrons (equation 27). For constant $R_i$ and $B$, the cooling-tail is $N(\gamma < \gamma_i) \sim \gamma^{-2}$ (Figure 2), and the corresponding low-energy photon slope is $\alpha = -3/2$. Before $t_{ci}$, the electron distribution has a cooling break above $\gamma_i$, across which the effective distribution steepens by unity and the spectral slope $\beta$ by 1/2. At $t < t_{ci}$, equation (13) with constant $R_i$ indicates a linear increase of the electron number and, hence, of the peak flux; at $t > t_{ci}$, equation (14) with $k_i$ = const and $x = 0$ (constant $B$) indicates a constant electron distribution nomalization, thus the constant spectral peak flux. **Both panels**: As the cooling-tail develops, it leads to a spectral widening (increasing WHM) and softening as shown by the decreasing $E_p$ (see also Figure 5) and by the decrease of the low-energy photon slope $\alpha$ from an initial asymptotic $\alpha = -2/3$ to the asymptotic value expected in each case, after the cooling-tail "fills" the observing window. At times later than shown, there is no significant spectral evolution because the electron distribution in the GRB observing window does not change. Consequently, pulse-**integrated** spectra are very close to the instantaneous spectrum at the latest epoch.

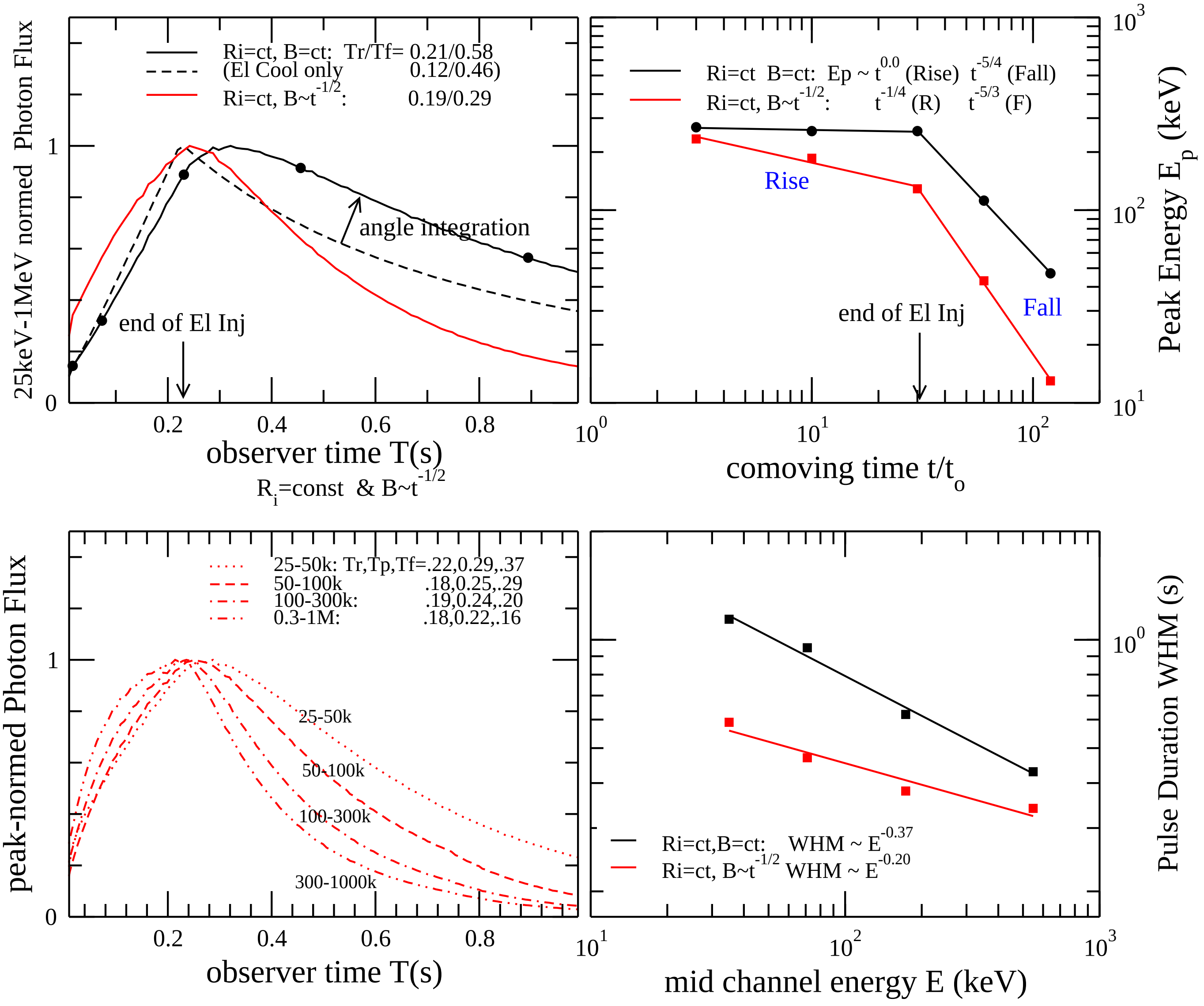

FIG. 5.— Pulse shape/duration and evolution of the peak-energy $E_p$ for **adiabatic** electron cooling. Parameters: typical energy of injected electrons $\gamma_i = 10^5$ (fixed), power-law index $p = 4$ above $\gamma_i$, source Lorentz factor $\Gamma = 100$. Injection rate $R_i$= const, magnetic field $B = 10$ G at $t = 10t_o$, with $t_o = 1$ s the initial comoving-frame age of the system. Electrons are injected until $t_i = 30\,t_o$, after that injection is switched off to obtain a pulse decay. The *pulse rise* corresponds to an increase in the number of electrons that radiate in the observing window. The *pulse decay* is due to electrons cooling below that window and/or to a decreasing $B$. **Left upper** panel: *Pulse shape depends on* $R_i(t)$ *and* $B(t)$. Those used here lead to pulse time-asymmetry ratios $T_r/T_f$ compatible to those observed, between 0.15 and 0.75 (Norris et al 1996). The pulse shape is also determined by the spread in photon arrival-time and energy due to the spherical curvature of the emitting surface, which delays the pulse peak, increases the pulse duration, and makes pulses more symmetric (dashed line shows the pulse shape resulting from electron cooling alone). Legend gives the flux rise-time $T_r$ from half-peak to peak and the fall-time $T_f$ from peak to half-peak. Dots indicate epochs $t = 3, 10, 30, 60, 120\,t_o$ (see right panel), and the conversion to observer time is $T = t/2\Gamma$. **Right upper** panel: *Evolution of the peak-energy* $E_p$ of the $\varepsilon F_\varepsilon$ spectrum. The power-law fits to the numerical $E_p$ shown here: $E_p = const$ before the pulse peak and $E_p \sim t^{-(1.3\div1.7)}$ after the peak, are close to expectations (equation 47). **Left lower** panel: *Pulses peak earlier and are shorter at higher energies.* Only the $R_i =$ const and $B \sim t^{-1/2}$ case is shown here, but results are similar for the other cases that yield pulses with correct shape. Legend gives the rise, peak, and fall times for four channels with increasing energy (25–50, 50–100, 100–300, 300–1000 keV, respectively) and indicates that: $i$) the average peak-time shift between adjacent channels is $\delta T_p$/WHM = -0.02/0.5 s, being compatible with the $\delta T_p \in (-0.1, 0.03)$ reported by Norris et al (1996) for pulses with WHM$\in$ $(0.15, 1.5)$ s; $ii$) pulses are more time-symmetric at higher energy, a potential contradiction with observations (Norris et al 1996, Lee et al 2000), which show an energy-independent pulse time-asymmetry. **Right lower** panel: *Pulses are shorter at higher energies.* The pulse width-at-half-maximum decreases with photon energy as WHM$\sim$ $E^{-(0.2\div0.4)}$, which is consistent with the measured WHM$\sim$ $E^{-0.4}$ (Fenimore et al 1995, Norris et al 1996).

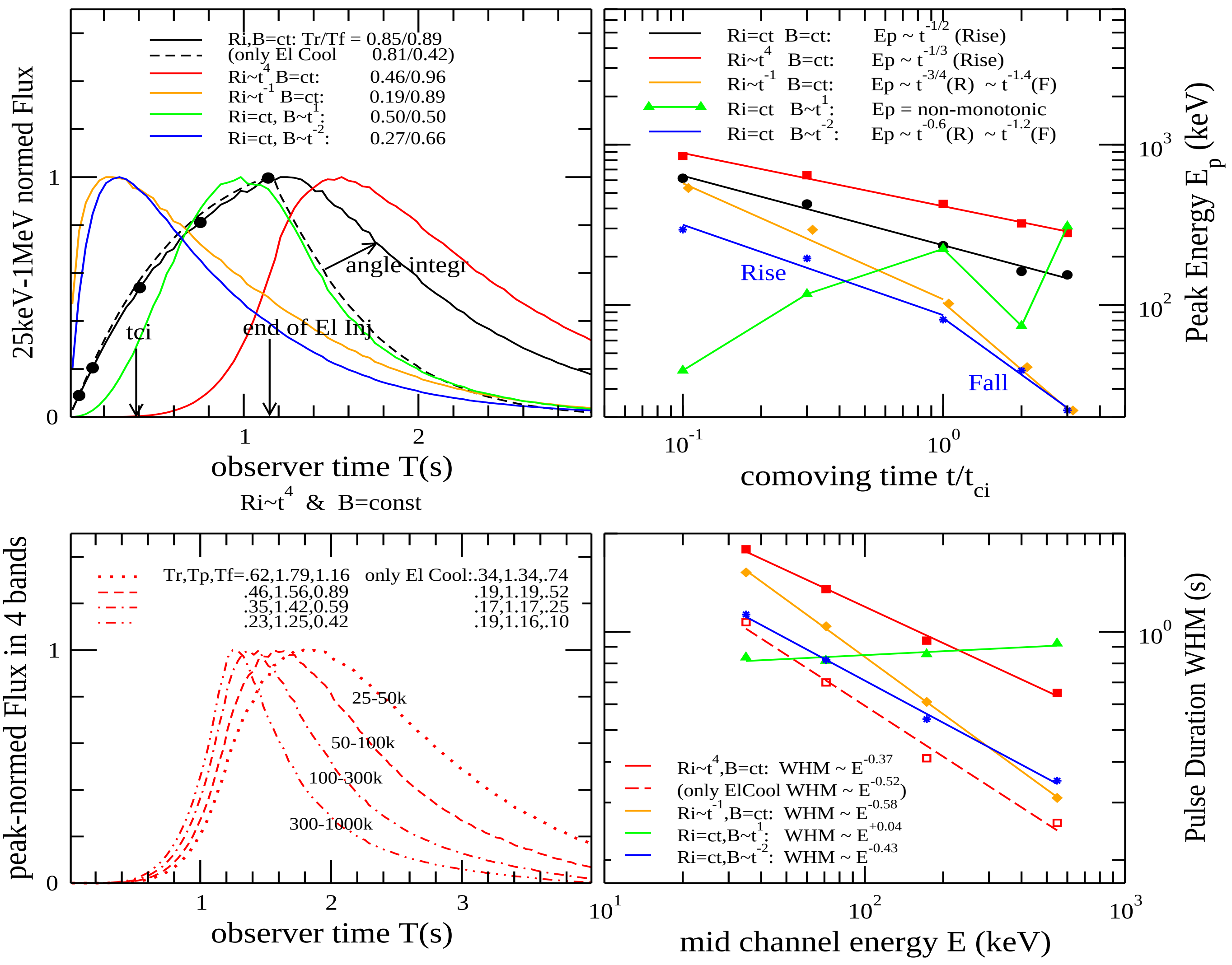

FIG. 6.— Pulse shape/duration and evolution of the peak-energy $E_p$ for **synchrotron**-dominated electron cooling. Parameters: $\gamma_i = 10^5$, $p = 3$, $\Gamma = 100$, $B = 10$ G at $t = t_{ci}$, thus $t_{ci} = 77$ s. Electrons are injected until $t_i = 3\,t_{ci}$, after that injection is switched off. **Left upper** panel: *Pulse shape dependence on* $R_i(t)$ *and* $B(t)$. For constant injection rate $R_i$ and magnetic field $B$, the pulse (black line) is almost time-symmetric; to obtain pulses that are as time-asymmetric as observed requires evolving $R_i(t)$ (red and orange lines) or $B(t)$ (green and blue lines). Legend gives the pulse rise and fall times $T_r$, $T_f$. The pulse shape resulting from electron cooling alone (without integrating emission over the source spherically-curved surface) is illustrated by the dashed line. Dots indicate epochs $t = 0.1, 0.3, 1, 2, 3\,t_{ci}$ (as in right panel). **Right upper** panel: *Evolution of the* $E_p$ *peak energy* of the $\varepsilon F_\varepsilon$ spectrum. For all but an increasing $B$, we obtain $E_p \sim T^{-(1/3 \div 3/4)}$ during the pulse rise and $E_p \sim T^{-1.3}$ after the pulse peak. **Left lower** panel: *Pulses peak earlier and are shorter at higher energies.* Only the $R_i \sim t^4$ and $B =$ const case is shown here, but results are similar for the other cases where $E_p$ decreases. The rise, peak, and fall times given in legend indicate that: $i)$ the average fractional peak-time shift between adjacent channels is $\delta T_p$/WHM = -0.15/1.0 s; somewhat larger than observed ($\delta T_p \in (-0.1, 0.03)$ for WHM$\in (0.15, 1.5)$ s) and clearly larger than for adiabatic cooling (Figure 5), $ii)$ pulse asymmetry factor is energy-independent ($T_r/T_f \simeq 0.55$ in all channels), consistent with observations. Electron cooling makes pulses peak earlier, be shorter, and more symmetric at higher energy, while the curvature of the emitting surface strenghtens the first feature, preserves the second, and supresses the last. **Right lower** panel: *Pulses are shorter at higher energies*, except for an increasing $B$ (which leads to a nearly constant $E_p$). The resulting WHM$\sim E^{-(1/3 \div 1/2)}$, for the models that yield a correct pulse time-asymmetry, is consistent with the measured WHM$\sim E^{-0.4}$.